\documentclass[longauth]{aa}  

\usepackage{graphicx,amssymb,amsmath,times,xcolor}
\usepackage{longtable}
\usepackage{bm} 
\usepackage{url}
\usepackage{subcaption}
\usepackage{graphicx} 

\usepackage[varg]{txfonts}
\usepackage[T1]{fontenc}
\usepackage{natbib}
\bibpunct{(}{)}{;}{a}{}{,} % to follow the A&A style
\usepackage[normalem]{ulem}

\begin{document}
   \title{Theoretical Pulsation Properties of RR Lyrae and Type II Cepheids in the JWST and Roman photometric systems}
%   \subtitle{Pulsation properties and relations}

\author{
M.~Marconi \inst{1}
\and R.~Molinaro \inst{1}
\and A.~Bhardwaj \inst{2}
\and I.~Musella \inst{1}
\and V.~Ripepi \inst{1}
\and E.~Trentin \inst{1}
\and G.~De Somma \inst{1,3}
\and T.~Sicignano \inst{4,5}
\and M.~Deka \inst{1}
\and M.~Gholami \inst{1}
\and E.~Luongo \inst{1,6}
\and M.~Dall'Ora \inst{1}
\and S.~Leccia \inst{1}
}

\institute{
 INAF-Osservatorio Astronomico di Capodimonte, Salita Moiariello 16, 80131 Napoli, Italy
\and Inter University Center for Astronomy and Astrophysics (IUCAA), Pune, India 
\and Istituto Nazionale di Fisica Nucleare (INFN) - Sez. di Napoli, Compl. Univ.di Monte S. Angelo, Edificio G, Via Cinthia, I-80126, Napoli, Italy
\and European Southern Observatory, Karl-Schwarzschild-Str. 2, 85748 Garching bei München, Germany
\and Scuola Superiore Meridionale, Largo S. Marcellino 10, 80138 Napoli, Italy
\and Università di Salerno, Dipartimento di Fisica “E.R. Caianiello”, Via Giovanni Paolo II 132, 84084 Fisciano (SA), Italy
%\and INAF-Osservatorio Astronomico d'Abruzzo, Via Maggini sn, 64100 Teramo, Italy
%\
}

   \date{Received XXXX; accepted XXXX}

% \abstract{}{}{}{}{}
% 5 {} token are mandatory
 
  \abstract
   % Context (optional)
   {RR Lyrae (RRL) stars and Type II Cepheids (T2Cs) are key standard candles and tracers of old stellar populations. The deep, wide-field capabilities of facilities such as the James Webb Space Telescope (JWST) and the Nancy Grace Roman Space Telescope will significantly expand their utility as a Population II alternative to Classical Cepheids for distance scale calibrations.}
  % Aims (mandatory)
   {We present theoretical pulsation properties—including light-curve morphologies, amplitudes, and multi-filter relations—for RRLs ($Z = 0.0001$--$0.02$, $M = 0.51$--$0.80\,M_{\odot}$) and T2Cs ($Z = 0.0001$--$0.01$, $M = 0.495$--$0.65\,M_{\odot}$) to enable detailed studies of Galactic and extragalactic populations.}
  % Methods (mandatory)
   {Using alreday available non-linear convective pulsation models, we converted their predicted bolometric light curves across various chemical compositions and evolutionary stages into the JWST and Roman photometric systems. From these, we derived multi-filter Period-Amplitude and Color-Color diagrams, alongside Period-Luminosity and Period-Wesenheit relations with explicit metallicity terms.}
  % Results (mandatory)
   {RRL relations display a significantly stronger zero-point metallicity dependence than those of T2Cs, despite sharing similar slopes. Consequently, T2Cs offer a distance scale indicator that is remarkably robust against metallicity uncertainties. As a preliminary test, we applied our metal-dependent Period-Wesenheit relations to a subset of JWST-observed RRLs in the WLM galaxy, deriving a distance modulus of $\mu = 24.85 \pm 0.03$~mag (assuming the literature average metallicity).}
  % Conclusions (mandatory)
   {The derived distance to WLM is consistent with most literature values, though slightly shorter than determinations based on Classical Cepheids and the Tip of the Red Giant Branch. These theoretical relations pave the way for future high-precision distance determinations and asteroseismic diagnostics with JWST and Roman.}
  
   {}


   \maketitle
   \nolinenumbers
%
%________________________________________________________________

\section{Introduction}
RR Lyrae (RRL) stars and Type II Cepheids (T2Cs) are well known standard candles and  tracers of the formation and evolution of old stellar populations. In the case of RRL, their use as distance indicators has traditionally relied upon the well-established luminosity–metallicity relation \citep[see, e.g.,][and references therein]{caputo00,dicrisci04,marconi2018impact,Muraveva18}  and a period–luminosity (PL) relation in the near-infrared (NIR) filters \citep[see, e.g.,][]{longmore1986rr,Bono01,Bono03,2004ApJS..154..633C,Dallora06,Coppola11,2015A&A...577A..99N,marconi2015,Muraveva15,marconi2022}. Moreover, accurate two-filter and three-filter Period-Wesenheit (PW) relations can be derived for these pulsators \citep[see, e.g.,][]{marconi2015}. While the physical basis for these relations is robust, the precise values of coefficients and the role of metallicity in the NIR PL and PW remain a subject of active debate \citep[see, e.g.,][]{Bono03,2009A&A...502..505B,marconi2015,Muraveva18,marconi2022,Bhardwaj23,2023ApJ...951..114Z,2024A&A...684A.176P}.

From the observational point of view, the metallicity of RR Lyrae stars can be derived either through high resolution spectroscopy \citep[see e.g.][and references therein]{Crestani21,Gilligan21,Dorazi24} or photometrically through the dependence on metal abundance of Fourier parameters \citep[see e.g.][and references therein]{JK96,Dekany21,Li23,Prakash26}.

Nonlinear convective pulsation models frequently predict a significant metal dependence for RRL variable stars \citep[see, e.g.,][]{marconi2015,marconi2022}, whereas some empirical studies suggest a milder or negligible effect. However, recent high-precision calibrations \citep[see, e.g.,][]{Bhardwaj23}  quite closely align with theoretical predictions for the metallicity term coefficient both in PL and PW relations of RRL.

T2Cs are significantly more luminous than RRL, typically by more than 0.2 dex, making them powerful distance indicators across diverse stellar environments. They have been identified in the Galactic Bulge \citep[e.g.,][]{matsunaga13,2017A&A...605A.100B, braga19}, in Galactic Globular Clusters \citep[GGCs,][]{matsunaga2006,ngeow22}, in the Galactic Halo \citep[][]{2022ApJ...927...89W,Ripepi2023} and in the Magellanic Clouds \citep[e.g.,][]{soszynski18,sicignano24}. Classified by their pulsation periods, T2Cs are traditionally divided into three sub-groups: BL Herculis stars ($P < 4$ days), W Virginis stars ($4 < P < 20$ days), and RV Tauri stars ($P > 20$ days). While foundational stellar evolution models for T2Cs were established by Gingold (1976, 1985), more recent investigations have focused primarily on the pulsation properties of the BL Herculis subclass \citep[e.g.,][]{dicriscienzo07,bono20,2021MNRAS.501..875D,2025A&A...695A..38D}. In \citet{Marconi2025A&A...703A.228M}, we expanded upon this theoretical landscape by computing the oscillation properties of post-Zero-Age Horizontal Branch (ZAHB) stars directly along their evolutionary tracks. On this basis, we presented—for the first time—a detailed pulsation analysis across three distinct metallicity values, bridging the gap between evolutionary theory and observed pulsation properties.
The next era of time-domain astronomy is defined by a shift toward high-cadence, deep-sky infrared and optical mapping. The Vera C. Rubin Observatory’s Legacy Survey of Space and Time \citep[LSST,][]{Ivezic2019} is expected to revolutionize our census of variable stars by providing a 10-year "movie" of the southern sky. Complementing this is the Nancy Grace Roman Space Telescope \citep[Roman,][]{Akeson2019Roman}, whose Wide Field Instrument will provide Hubble-like resolution over a field of view 100 times larger, focusing on the Galactic Bulge and high-latitude fields. Together with the JWST, these missions will extend our reach to variable stars in distant, obscured environments that were previously inaccessible, allowing us to use them as precision distance markers across the Local Universe.
 
Fully exploiting observational data from these missions for distance scale estimation depends heavily on developing a robust theoretical framework for pulsating stars in the corresponding photometric filters. In \citet{marconi2022}, \citet{DeSomma24} and \citet{Marconi2025A&A...703A.228M}, we provided a theoretical scenario for the Rubin-LSST for RRL, Classical Cepheids (CCs) and T2Cs, respectively. Here we present the predicted pulsation properties of RRL and T2Cs in the unique filter systems of Roman and JWST. A parallel investigation for CCs has been recently published \citep{DeSomma26}.

Unlike traditional optical filters (like Johnson V or Gaia G), the near-infrared (NIR) pass bands of these new observatories significantly reduce the impact of interstellar dust on the PL relations. Higher spatial resolution and deep photometric time-series from these facilities will provide observational properties of
RRL/T2Cs in previously inaccessible stellar environment. However, to fully exploit these time-series databases,  nonlinear convective pulsation models predicting the light and color curves in these specific NIR filters are needed. 

In Section 2, we present an atlas of predicted light curves for fundamental (F) and first-overtone (FO) RRL and T2Cs pulsators, along with the corresponding Bailey diagrams (amplitude vs. period) tailored to Roman and JWST filters. In Section 3, we investigate the behaviour of these models in various color–color planes, identifying specific indices that could serve as reddening indicators. In Section 4, we derive new theoretical Period–Luminosity(–Metallicity) ($PLZ$) and Period–Wesenheit(-Metallicity) ($PWZ$) relations in the JWST and Roman bands for the two classes of pulsating stars. Finally, Section 5 addresses the conclusions and future perspectives.
%__________________________________________________________________

\section{The model light curves in the Roman and JWST filters}

\begin{figure*}
  \centering
  \begin{subfigure}[b]{0.77\linewidth}
    \centering
    \includegraphics[trim=0cm 3.0cm 0.0cm 0cm, clip, width=1.0\linewidth]{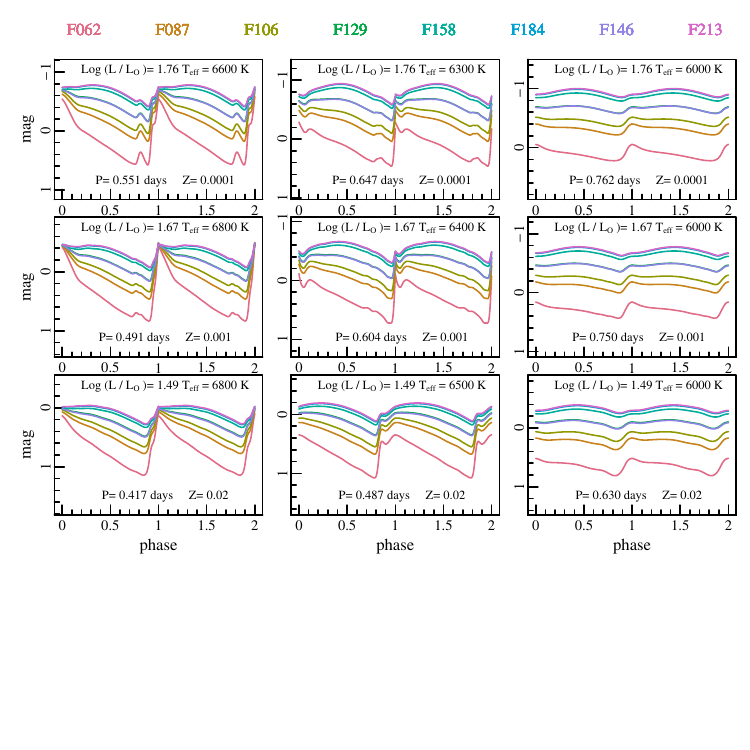}
    \caption{Roman photometric system}\label{fig-roman_lightCurves_example_rrl_F_roman}
  \end{subfigure}
  
  \vspace{0.1cm} % Spaziatura verticale opzionale tra le due righe
  
  \begin{subfigure}[b]{0.77\linewidth}
    \centering
    \includegraphics[trim=0cm 3.0cm 0.0cm 0cm, clip, width=1.0\linewidth]{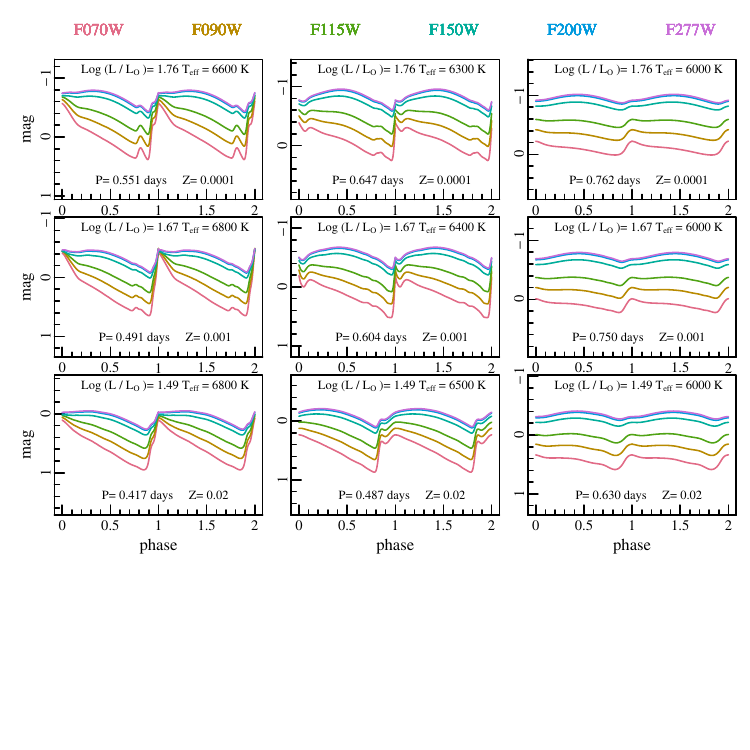}
    \caption{JWST photometric system}\label{fig-jwst_lightCurves_example_rrl_F_jwst}
  \end{subfigure}
\vspace{-0.1cm}
  \caption{Examples of theoretical F-mode RR Lyrae light curves as obtained from nonlinear convective pulsation models transformed into the Roman (top) and JWST (bottom) photometric systems. Each panel shows the predicted magnitude variation as a function of the pulsation phase in different filters, color-coded as labeled on top. From left to right, panels show models with decreasing effective temperatures within the instability strip. Moving from top to bottom within each set, metallicity increases from $Z = 0.0001$ to $Z = 0.02$. The adopted stellar masses are $0.80\,M_{\odot}$ (top), $0.64\,M_{\odot}$ (middle), and $0.54\,M_{\odot}$ (bottom). Luminosity levels correspond to ZAHB values for each metallicity. Nonlinear pulsation periods are labeled in each panel.}\label{fig-lightCurves_combined}
\end{figure*}

\begin{figure*}
  \centering
  \begin{subfigure}[b]{0.77\linewidth}
    \centering
  \includegraphics[trim=0cm 3.0cm 0.0cm 0cm, clip, width=1.0\linewidth]{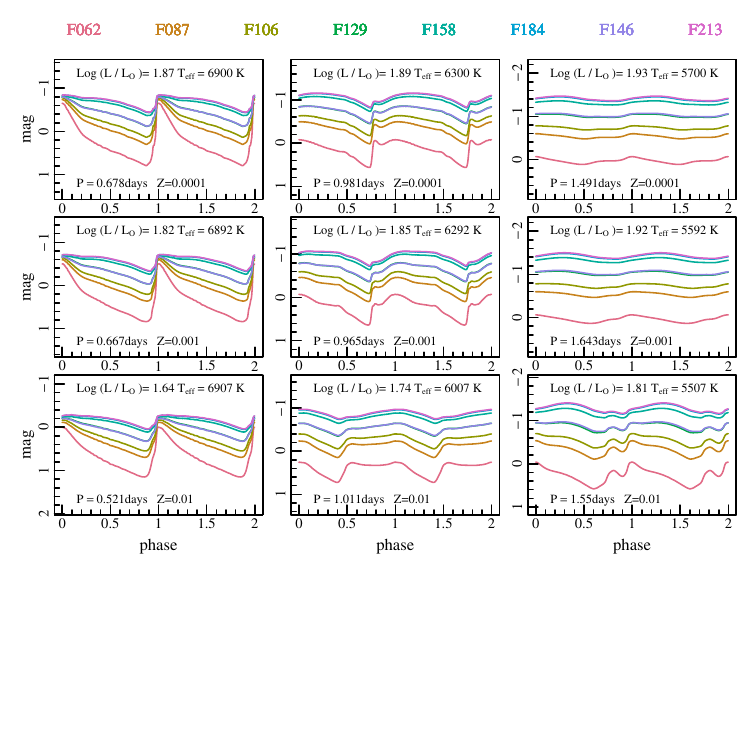}
    \caption{Roman photometric system}\label{fig-roman_lightCurves_example_rrl_F_roman}
  \end{subfigure}
  
  \vspace{0.1cm} % Spaziatura verticale opzionale tra le due righe
  
  \begin{subfigure}[b]{0.77\linewidth}
    \centering
  \includegraphics[trim=0cm 3.0cm 0.0cm 0cm, clip, width=1.0\linewidth]{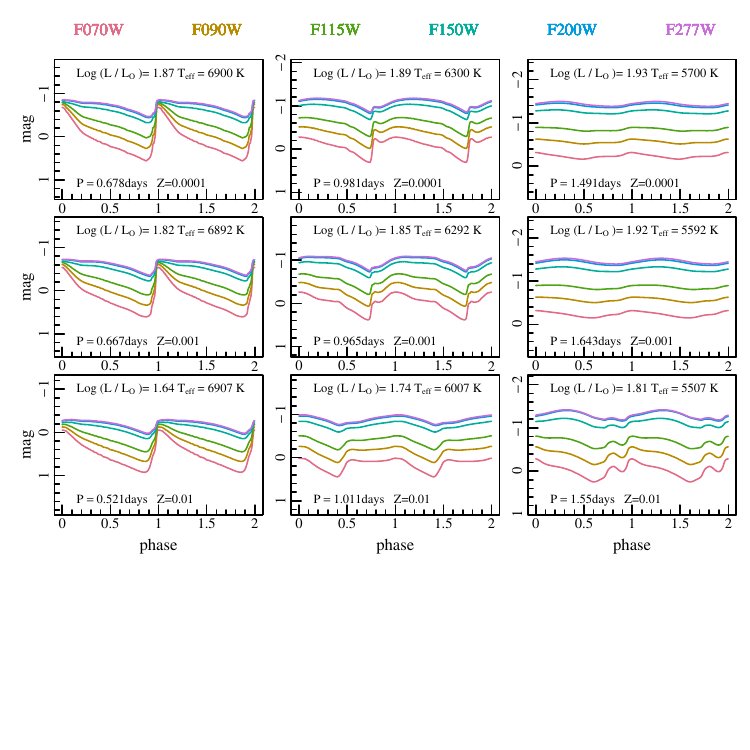}
    \caption{JWST photometric system}\label{fig-jwst_lightCurves_example_rrl_F_jwst}
  \end{subfigure}
\vspace{-0.1cm}
  \caption{Examples of theoretical T2Cs light curves as obtained from nonlinear convective pulsation models transformed into the Roman (top) and JWST (bottom) photometric systems. The stellar parameters labelled in the various panels correspond, for each metallicity and mass, to values selected along the predicted evolutionary tracks \citep[see][and references therein, for details]{Marconi2025A&A...703A.228M}.
 In particular, the adopted stellar masses are 0.65, 0.60, 0.55 $M_{\odot}$ for Z=0.0001 (top panels), 0.001 (middle panels) and 0.01 (bottom panels), respectively. For each $Z$ and mass the luminosity/effective temperature increases/decreases from left to right.}\label{T2Cfig-lightCurves_combined}
\end{figure*}

We took into account the nonlinear convective pulsation models already presented in our previous papers for RRL \citep{marconi2015,marconi2018impact,marconi2022} and T2Cs \citep{Marconi2025A&A...703A.228M}. These models, in the case of RRL pulsators cover an extensive range and a detailed grid of stellar masses, luminosities and effective temperature for each selected chemical composition \citep[see][for details]{marconi2015}, while in the case of T2Cs specific combinations of mass, luminosity and effective temperature are selected along the predicted evolutionary tracks \citep[see][for details]{Marconi2025A&A...703A.228M}. In both cases a standard mixing length parameter of 1.5 was assumed in the adopted nonlocal time-dependent treatment of convection, to close the nonlinear equation system. The bolometric light curves obtained from these models have been transformed into the 
Roman and JWST bands by adopting the procedure described by \citet{Chen2019}.
In Figure \ref{fig-lightCurves_combined}  we show an example of the produced theoretical light curves in the Roman and JWST bands for F-mode RRL. In each plot, from top to bottom panels, the adopted metallicity increases from $Z=0.0001$ to $Z=0.02$ (see labels), whereas from left to right, the light curves of  models close to the blue edge, in the middle of the instability strip and close to the red edge are shown.
The adopted metallicity, luminosity, effective temperature, together with the inferred pulsation period are labelled in each panel. We note that, for each metallicity, the adopted stellar mass and the labelled luminosity correspond to the predicted ZAHB values. As expected we find progressively smaller amplitudes and smoother variations toward longer wavelengths.
A similar plot but for FO-mode RRL models is shown in Figure A.1 in Appendix A.
In the case of T2Cs, the predicted light curves in the Roman and JWST filters are shown in Figure \ref{T2Cfig-lightCurves_combined}. Only F-mode pulsators are considered in this case, given the very small number of predicted FO pulsators. The stellar parameters labelled in the various panels correspond, for each metallicity and mass, to values along the predicted evolutionary tracks \citep[see][and references therein, for details]{Marconi2025A&A...703A.228M}.
In particular, the adopted stellar masses are 0.65, 0.60, 0.55 $M_{\odot}$ for Z=0.0001, 0.001 and 0.01, respectively, and for each metallicity and mass the luminosity/effective temperature increases/decreases from left to right. As expected, and similarly to what was noted for RRL model curves, the pulsation amplitude of predicted T2C light curves decreases as the wavelength increases.

%   \begin{figure*}
%    \centering
%   \includegraphics[trim=0cm 3.0cm 0.0cm 0cm, clip, width=1.0\linewidth]{TypeII_curves_pro_Berna2026_roman.pdf}
% \caption{Model light curves for F-mode T2Cs in the Roman filters. The stellar parameters labelled in the various panels correspond, for each metallicity and mass, to values selected along the predicted evolutionary tracks \citep[see][and references therein, for details]{Marconi2025A&A...703A.228M}.
% In particular, the adopted stellar masses are 0.65, 0.60, 0.55 $M_{\odot}$ for Z=0.0001 (top panels), 0.001 (middle panels) and 0.01 (bottom panels), respectively. For each metallicity and mass the luminosity/effective temperature increases/decreases from left to right (see labels).}\label{fig-TypeII_curves_pro_Berna2026_roman}%
%     \end{figure*}

%   \begin{figure*}
%    \centering
%   \includegraphics[trim=0cm 3.0cm 0.0cm 0cm, clip, width=1.0\linewidth]{TypeII_curves_pro_Berna2026_jwst.pdf}
% \caption{Same as Fig.~\ref{fig-TypeII_curves_pro_Berna2026_roman} but for JWST bands}\label{fig-TypeII_curves_pro_Berna2026_jwst}%
%     \end{figure*}

\subsection{The period-amplitude diagrams}

\begin{figure*}
   \centering
  \includegraphics[trim=0cm 0cm 0.0cm 0cm, clip, width=0.7\linewidth]{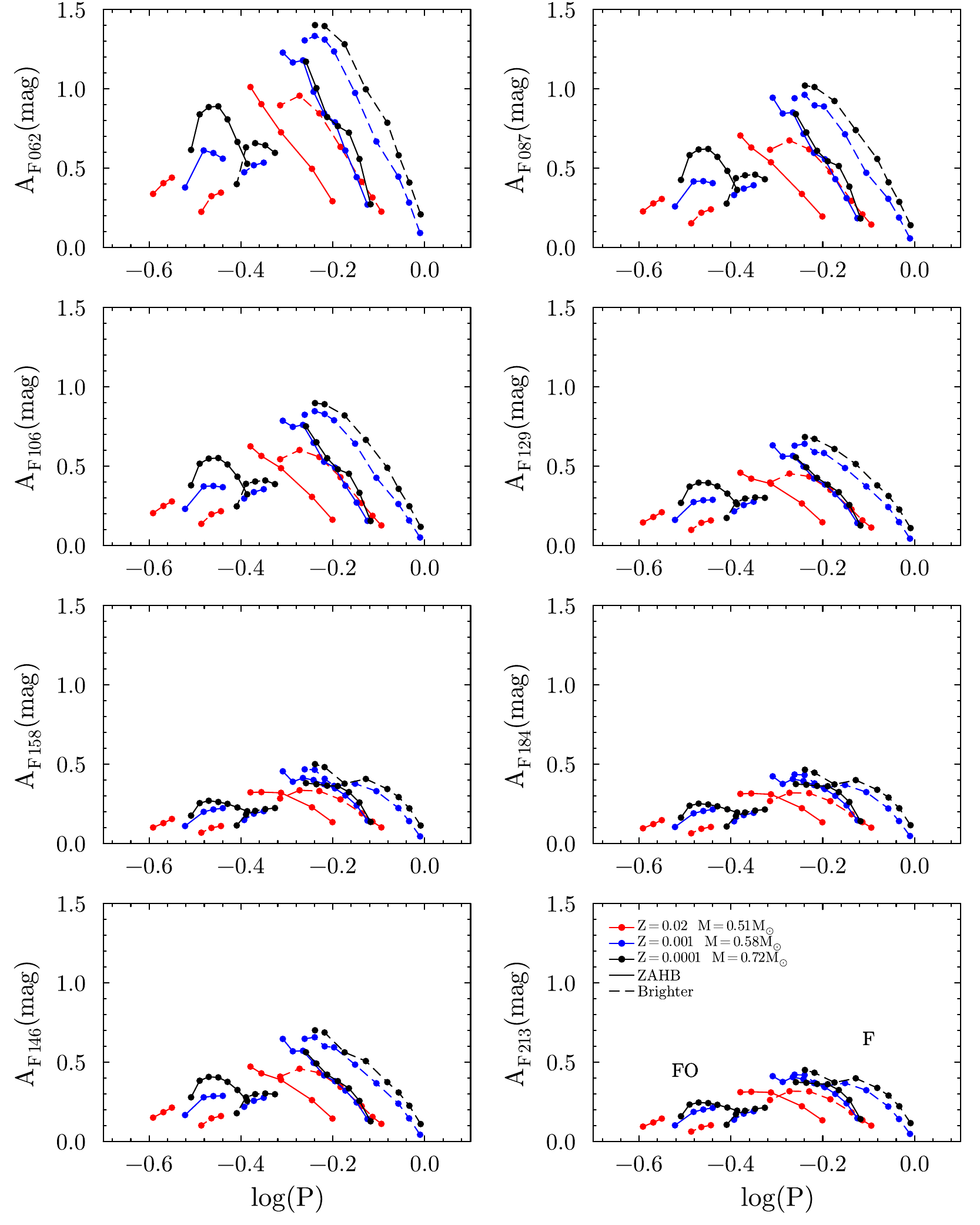}
\caption{The period-amplitude diagram for RRL models in the Roman filters for the labelled metallicities, masses and luminosity levels. Both FO- and F-mode models are plotted. Models corresponding to the ZAHB luminosity level (solid lines)  and to a brighter luminosity by 0.1 dex (dashed lines) are shown for each metallicity.}\label{fig-bailey_zy_variable_rrlyrae_roman}%
    \end{figure*}

\begin{figure*}
   \centering
  \includegraphics[trim=0cm 0cm 0.0cm 0cm, clip, width=0.7\linewidth]{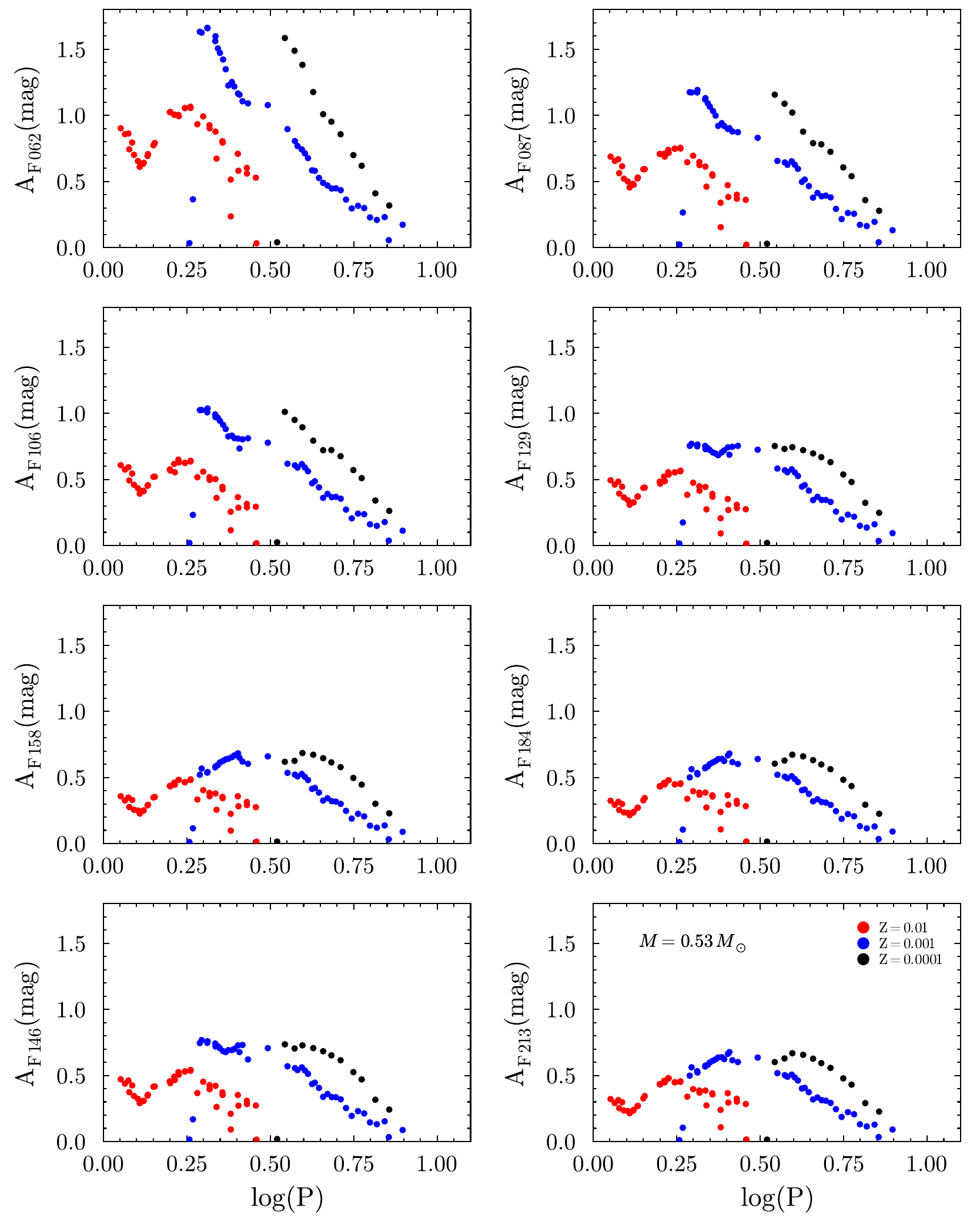}
\caption{The period-amplitude diagram for T2C models in the Roman filters. For each metallicity and mass (see labels) models have been selected along the evolutionary tracks \citep[see][for details]{Marconi2025A&A...703A.228M}.}\label{fig-bailey_zy_variable_typeII_roman_M0.53}%
    \end{figure*}

From the above-presented light curves we derived the pulsation amplitudes in the various photometric filters. In Figures \ref{fig-bailey_zy_variable_rrlyrae_roman} and \ref{fig-bailey_zy_variable_typeII_roman_M0.53} we show the inferred period-amplitude diagrams for F- and FO-mode RRL and F-mode T2C models, respectively, in the Roman bands for the labelled metallicities and masses.
In the case of RRL, models corresponding to the ZAHB luminosity level and to a brighter luminosity by 0.1 dex  are selected for each metallicity. In contrast, for T2Cs, models were selected along the evolutionary tracks for each mass and metallicity.
Similar plots but in the JWST filters are shown in Appendix B.
As already found in other photometric filters, the period-amplitude diagram in the Roman and JWST filters shows a decrease of periods and amplitudes as the metallicity increases and the already noted decrease of amplitudes as the wavelength increases. These occurrences hold both for RRL and T2C models.

The decrease of pulsation amplitude as the wavelength increases is accurately predicted by the adopted nonlinear convective pulsation models that are able to reproduce the dependence of  the brightness variation on how sensitive the Planck function is to temperature changes at a given wavelength. Indeed optical bands reside on the Wien tail/steeper slope of the Planck function, making flux variations highly sensitive to temperature fluctuations. As you move to infrared bands the flux dependence flattens toward the Rayleigh-Jeans regime, causing a drop in amplitude and the disappearance of secondary features \citep[see, e.g., discussion in][]{BMS99}. This is the reason why observationally and theoretically the light curves become smoother and with lower amplitude as the wavelength increases.

\section{The color-color diagrams}

\begin{figure*}
   \centering
  \includegraphics[trim=0cm 30cm 9.0cm 0cm, clip, width=0.7\linewidth]{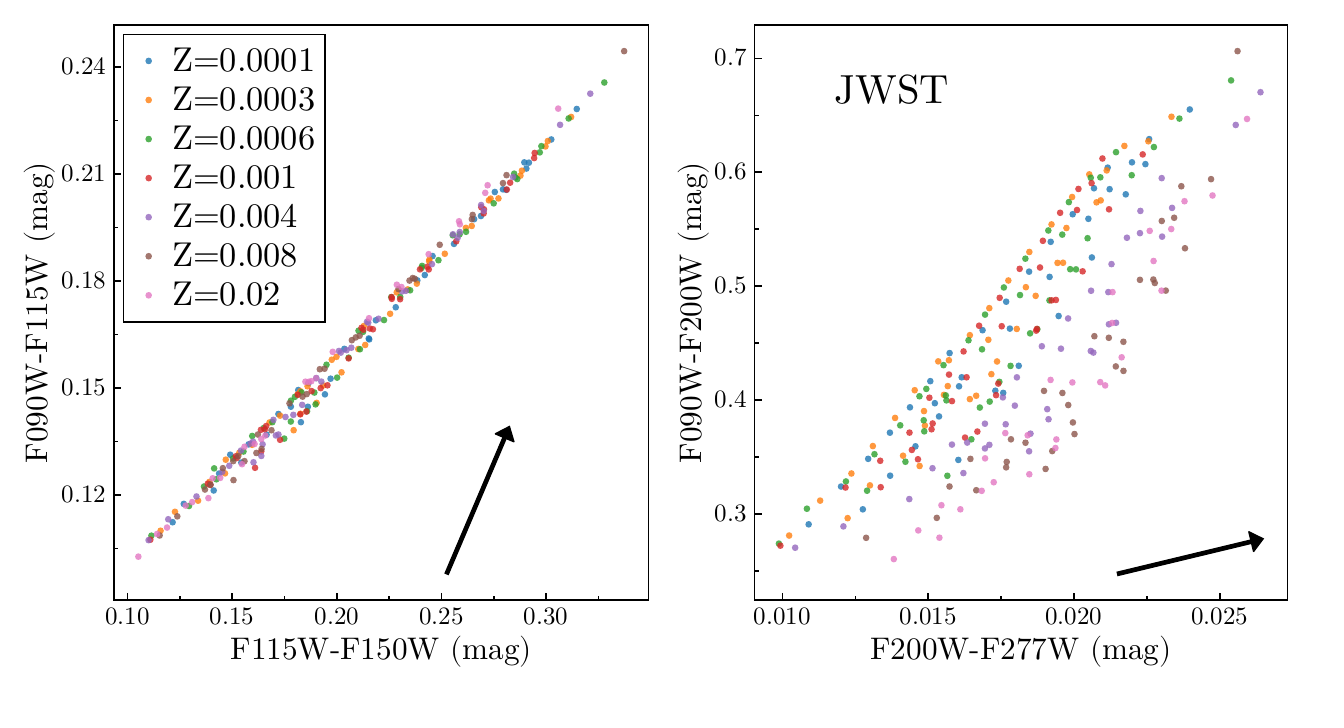}
\caption{Theoretical color-color diagrams for RR Lyrae variables in the James Webb Space Telescope (JWST) NIRCam photometric system. The left panel shows the [F090W-F115W] color index on the vertical axis against the [F115W-F150W] color index on the horizontal axis. The right panel displays the [F090W-F200W] color index versus [F200W-F277W]. The plotted points are derived from the mean magnitudes extracted from our theoretical light curves. Symbols of different colors indicate models computed at varying metallicity levels, ranging from Z=0.0001 to Z=0.02. The black arrows  indicate the direction of the reddening vectors for each specific JWST band combination.}\label{fig-rrlyrae_color_color_diagrams_jwst}
    \end{figure*}

\begin{figure*}
   \centering
  \includegraphics[trim=0cm 30cm 9.0cm 0cm, clip, width=0.7\linewidth]{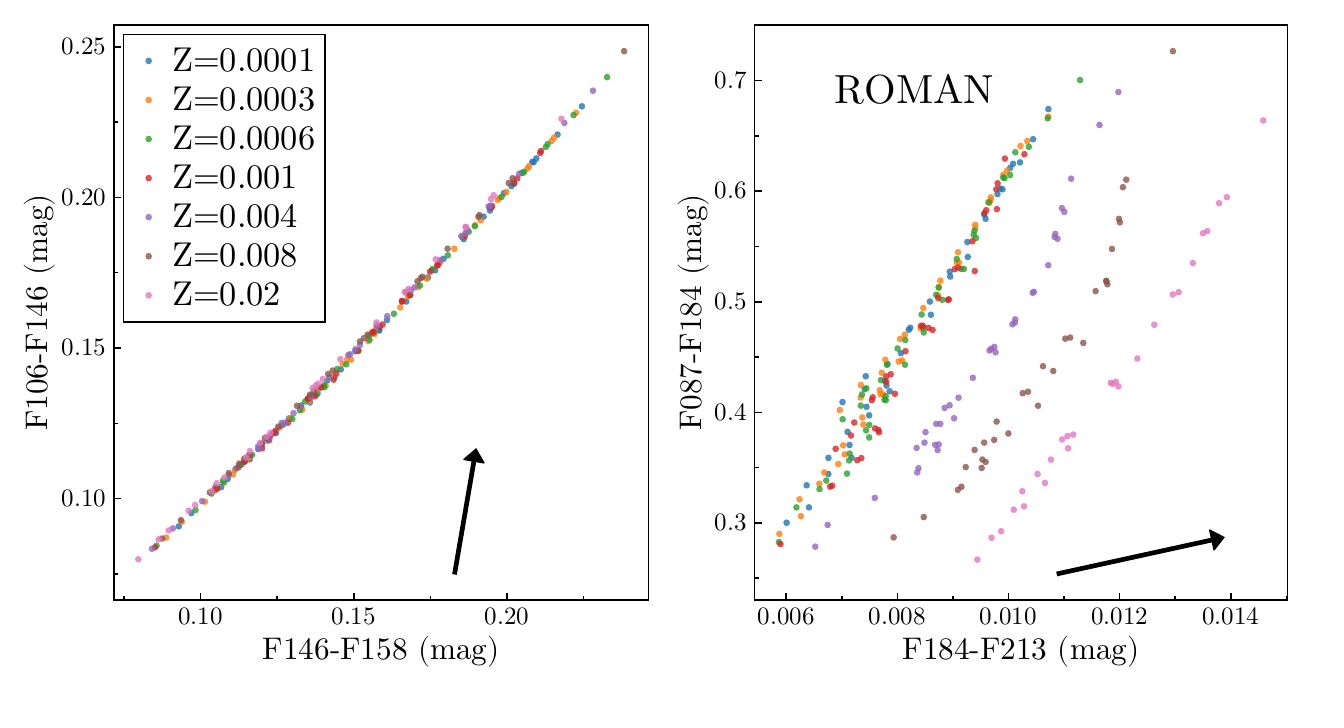}
\caption{Theoretical color-color diagrams for RRL variables in the Roman photometric system. The left panel shows the F106-F146 color index on the vertical axis against the F146-F158 color index on the horizontal axis. The right panel displays the F087-F184 color index versus F184-F213. The plotted points are derived from the mean magnitudes calculated from our theoretical light curves. Symbols are color-coded by metallicity, ranging from Z=0.0001 to Z=0.02. The black arrows  indicate the direction of the reddening vectors for each specific Roman band combination.
}\label{fig-rrlyrae_color_color_diagrams_roman}%
    \end{figure*}

As already discussed in previous theoretical analysis \citep[see, e.g.,][and references therein]{marconi2022} color-color plots can be used to constrain the reddening or stellar properties such as metallicity.
We explored several combinations of colors in the JWST and Roman bands. In Figure \ref{fig-rrlyrae_color_color_diagrams_jwst} we show RRL models computed for the labelled metallicity values in the (F090W-F115W) versus (F115W-F150W) and (F090W-F200W) versus (F200W-F277W) planes.
As for Roman bands, while Figure ~\ref{fig-rrlyrae_color_color_diagrams_roman} illustrates a couple of representative cases obtained by combining three photometric bands (e.g., F106, F146, and F158 in the left panel, and F087, F184, and F213 in the right panel), we systematically explored a broader range of color-color combinations, including multi-band diagnostic spaces leveraging four filters. Particularly striking is the behaviour of the models in the right panel of Figure ~\ref{fig-rrlyrae_color_color_diagrams_roman}. In this specific color-color plane (F087-F184 versus F184-F213), the sequences corresponding to different metallicities are clearly separated, even though the range on the X-axis covers only a few hundredths of a magnitude (ranging from approximately 0.006 to 0.014 mag). This occurs,  but in a less impactful way also in the JWST combination used for the right panel of Figure \ref{fig-rrlyrae_color_color_diagrams_jwst}.
 In all JWST and Roman color-color plots, the slope of the reddening vector differs, albeit to varying degrees, from the slope of model sequences, thus suggesting the possibility of constraining individual reddening values of investigated pulsators through the comparison between theory and observations in these diagrams. 
 We find a negligible metallicity dependence. Moreover, in the right panels of both figures the selected color pair provides theoretical patterns well aligned with the reddening vector, whereas in the left panels the displayed color-color patterns show a different slope suggesting the possibility of using these diagrams to infer individual reddening values of observed T2Cs.
 We emphasize that all the other explored band combinations, along with the exact slopes of the reddening vectors in the various planes, both for RRL and T2C models, are available upon request to the authors.

        \begin{figure*}
   \centering
  \includegraphics[trim=0cm 30cm 10.0cm 0cm, clip, width=0.7\linewidth]{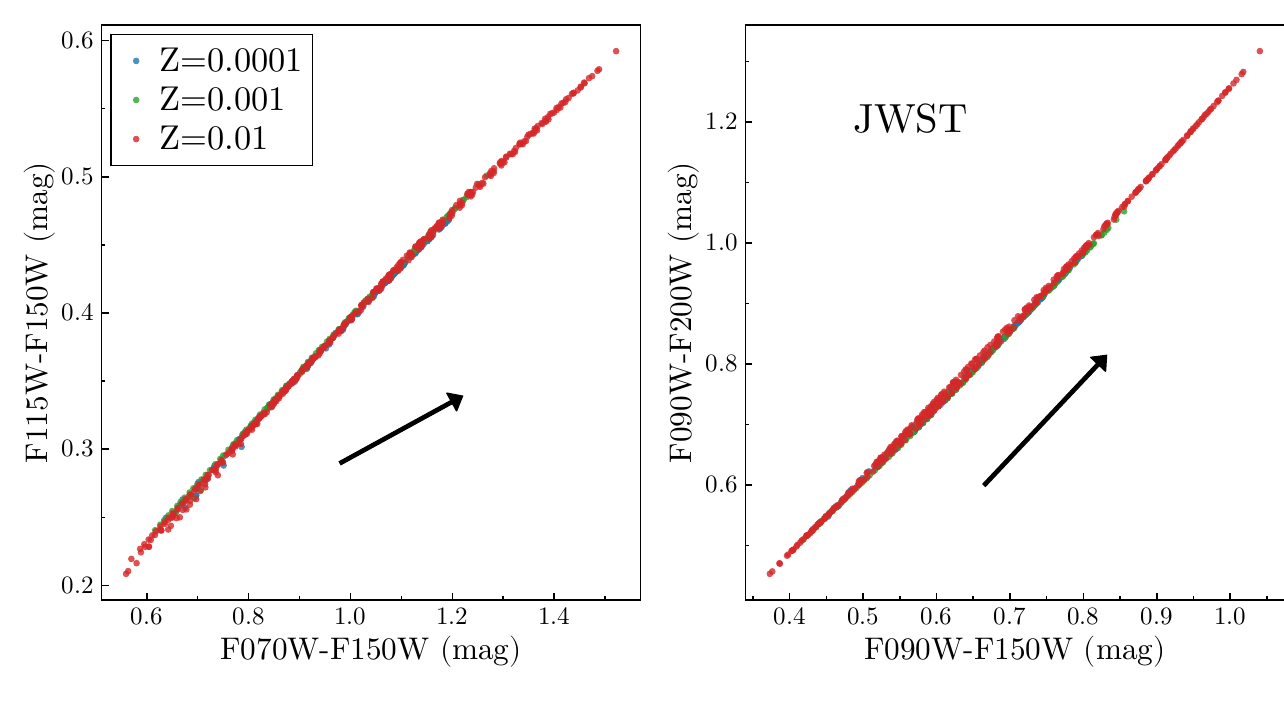}
\caption{Theoretical color-color diagrams for T2C models in the JWST photometric system. The left panel shows the F115W-F150W color index versus the F070W-F150W color index on the horizontal axis. The right panel displays the F090W-F200W color index versus the F090W-F150W color index on the horizontal axis. The plotted points are derived from the mean magnitudes calculated from our theoretical light curves. Symbols of different colors indicate models computed at varying metallicity levels, ranging from Z=0.0001 to Z=0.01. The black arrows indicate the direction of the reddening vectors for each specific JWST band combination.}\label{fig-type2_color_color_diagrams_jwst_user_ids}
\end{figure*}

\begin{figure*}
   \centering
  \includegraphics[trim=0cm 30cm 10.0cm 0cm, clip, width=0.7\linewidth]{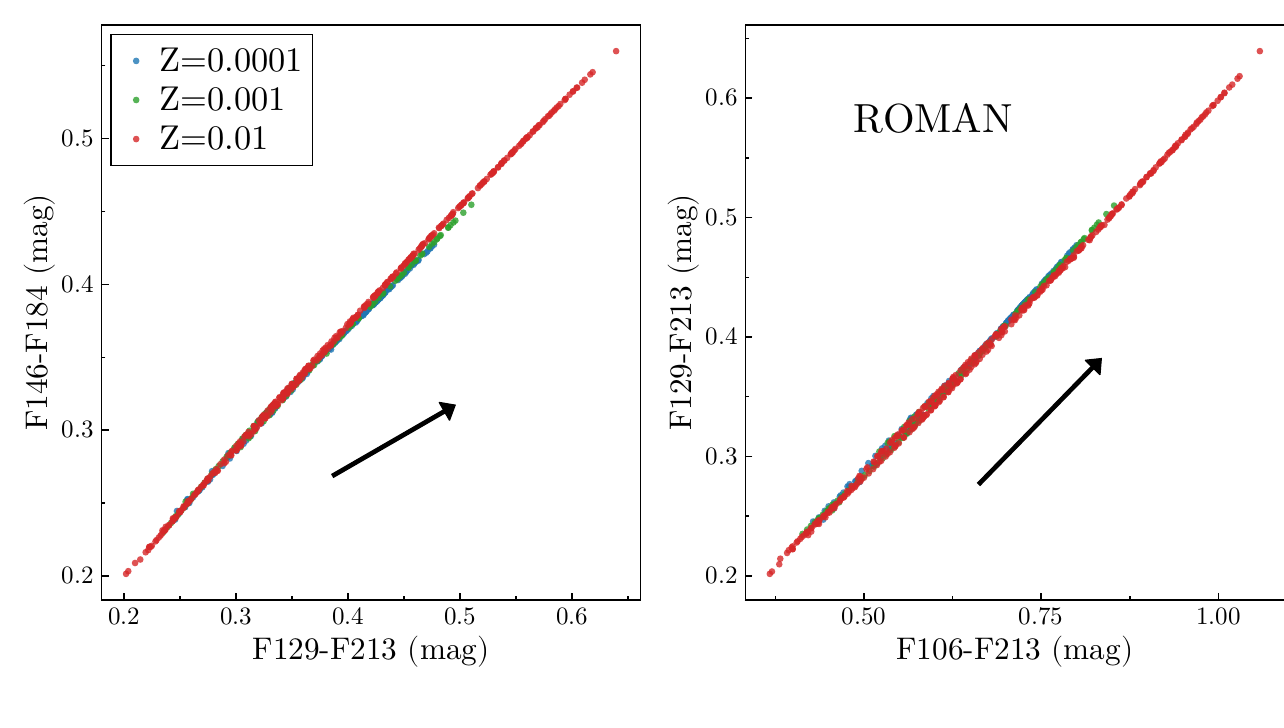}
\caption{Theoretical color-color diagrams for Type II Cepheid models in the Roman photometric system. The left panel shows the F146-F184 color index versus  the F129-F213 color index on the horizontal axis. The right panel displays the F129-F213 color index versus the F106-F213 color index on the horizontal axis. The plotted points are derived from the mean magnitudes calculated from our theoretical light curves. Symbols of different colors indicate models computed at varying metallicity levels, ranging from Z=0.0001 to Z=0.01. The black arrows indicate the direction of the reddening vectors for each specific Roman band combination.}\label{fig-type2_color_color_diagrams_roman_user_ids}
\end{figure*}

\section{The metal-dependent Period-Luminosity and Period-Wesenheit relations}
\begin{figure*}
   \centering
  \includegraphics[trim=0cm 0cm 0.0cm 0cm, clip, width=1.05\linewidth]{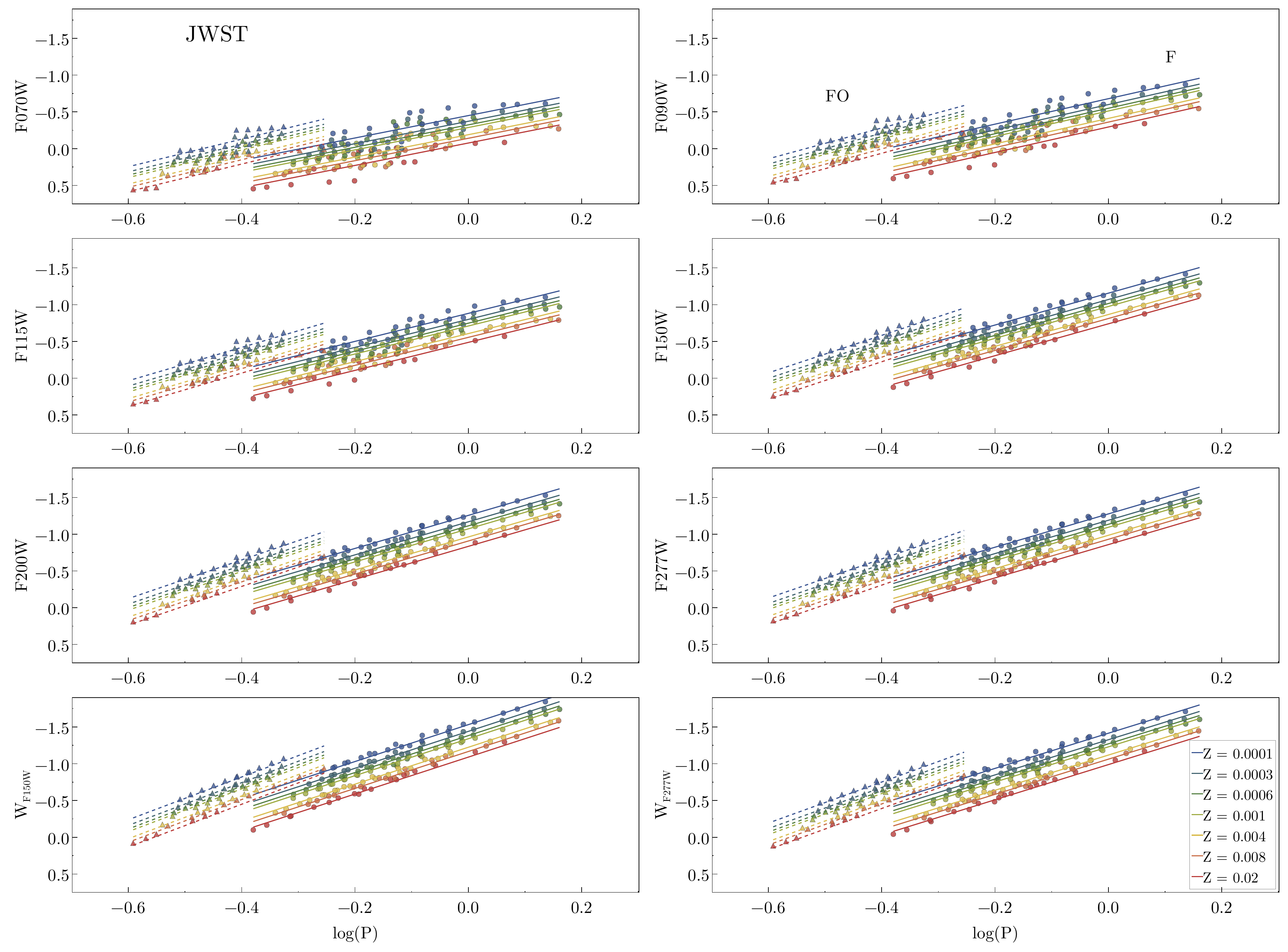}
\caption{Multi-filter PL and PW relations for the predicted JWST filters ($\rm F070W$, $\rm F090W$, $\rm F115W$, $\rm F150W$, $\rm F200W$, $\rm F277W$, $\rm W_{F150W}$, and $\rm W_{F277W}$) for RRL. Filled circles and triangles represent fundamental and first-overtone models, respectively. Data points are color-coded according to their metallicity Z. The solid lines show the best-fit analytical relations derived for the fundamental mode at fixed metallicity levels (for each $Z$, see legend), while dashed lines illustrate the corresponding fits for first-overtone models.}\label{fig-pl_feh_rrlyrae_jwst}%
    \end{figure*}

\begin{figure*}
   \centering
  \includegraphics[trim=0cm 0cm 0.0cm 0cm, clip, width=1.05\linewidth]{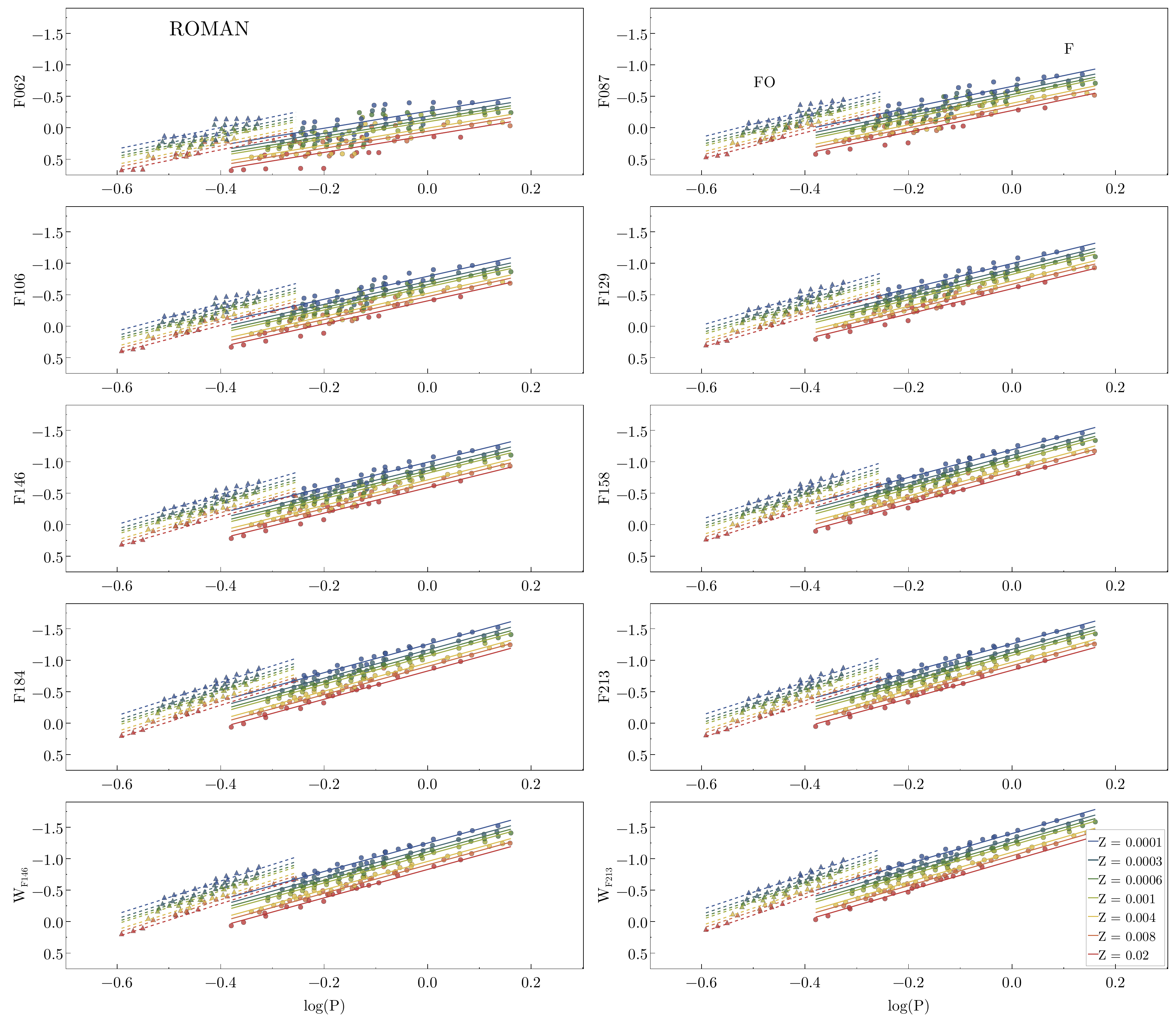}
\caption{Same as Fig.~\ref{fig-pl_feh_rrlyrae_jwst} but for the Roman photometric system.}\label{fig-pl_feh_rrlyrae_roman}%
    \end{figure*}

\vspace{-0.1cm}
To derive the predicted PLZ and PWZ relations, we adopted the model mean magnitudes and Wesenheit functions  in the various bands of both photometric systems, the corresponding pulsation periods, and the [Fe/H] values calculated from the heavy elements mass  abundance $Z$ of each model\footnote{In particular, we used the relation $\rm [Fe/H] = \log{(Z/Z_\odot)}$, where $\rm Z_\odot=0.02$ \citep{buldgen2024A&A...681A..57B}}. The results obtained for the two photometric systems analyzed in this work are detailed in Appendix C, where we report the coefficients of the selected metal-dependent PLZ and PWZ relations both in the JWST and Roman bands. In the case of RRL models, for each filter or filter pair, we derive the coefficients of the F, FO and (similarly to what was presented in our previous papers \citep[see e.g.][]{marconi2015,marconi2022}) fundamentalized relations. In the case of T2C models, the F-mode relations are reported.
 Figures \ref{fig-pl_feh_rrlyrae_jwst} and 
\ref{fig-pl_feh_rrlyrae_roman}  show the inferred relations for F and FO RRL models for the labelled metal abundances. The plotted symbols correspond to the individual pulsation models whereas the lines are the linear regression fits when replacing $[Fe/H]$ with the corresponding metal abundances.
The same relations but combining F and fundamentalized\footnote{FO models are fundamentalized by adding 0.127 to their $\log{P}$, as already discussed in \citet{marconi2015}.} FO RRL models are shown as global fits in Appendix D. 

We note that the observed deviations of some points from the linear regressions  arises because we plotted the overall relations (derived from the full model set) while adjusting only the metallicity term to match the value corresponding to each specific model subset. Consequently, the individual lines are not independent fits to each metallicity, but rather represent the global relation evaluated at different metallicities.

\begin{figure*}
   \centering
  \includegraphics[trim=0cm 0cm 0.0cm 0cm, clip, width=1.05\linewidth]{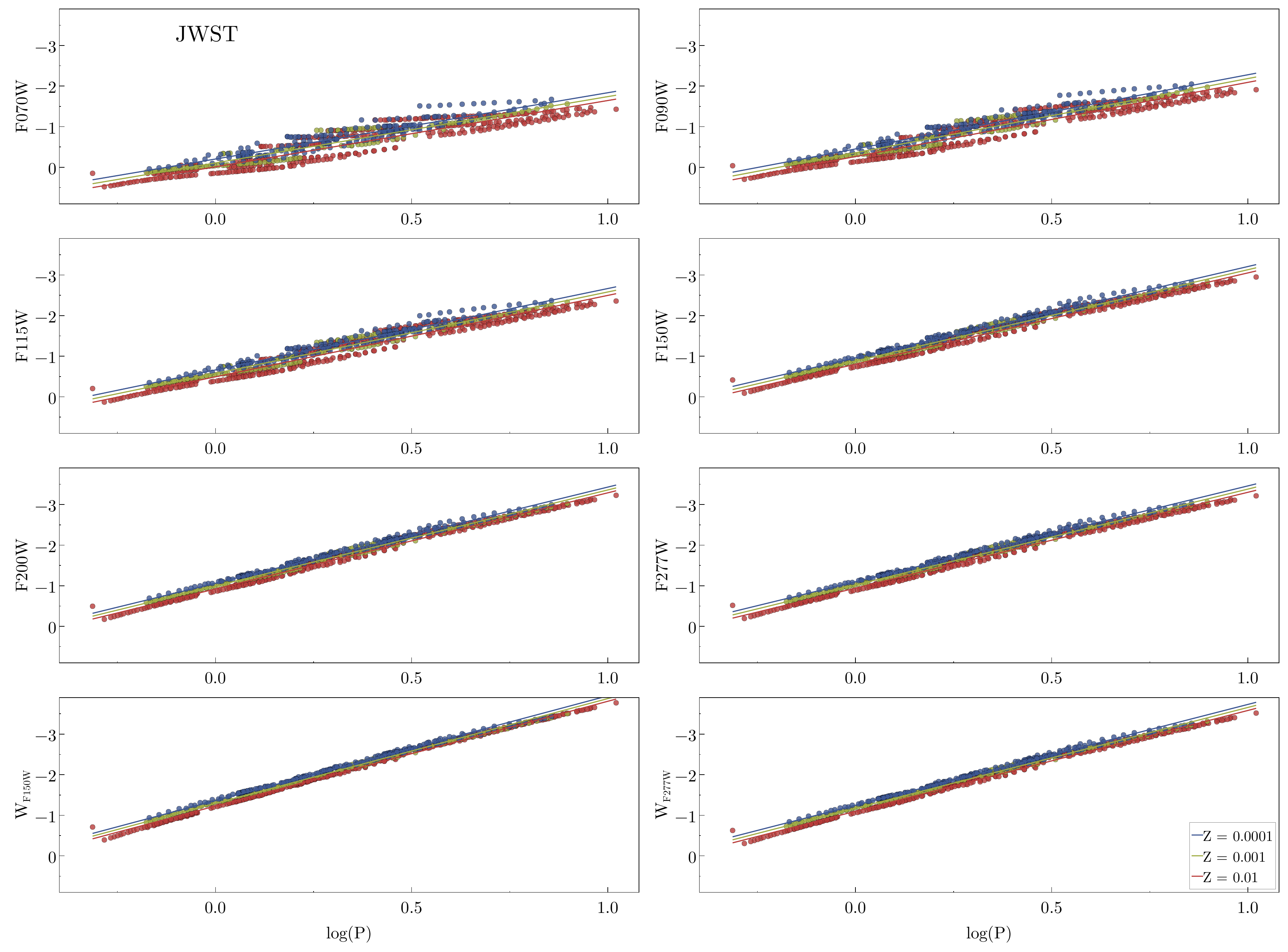}
\caption{Theoretical PL and PW relations for Type II Cepheid models in the JWST system. Panels from top to bottom show the relations for the individual filters (F062 through F213) and the Wesenheit magnitudes ($W_{\rm F146}$ and $W_{\rm F213}$, bottom row) as a function of the logarithm of the pulsation period ($\log P$, with $P$ in days). Individual models are color-coded according to their metallicity: $Z = 0.0001$ (blue), $Z = 0.001$ (green), and $Z = 0.01$ (red). Solid lines of corresponding colors represent the best-fit linear relations at each fixed chemical composition. The progressive reduction in dispersion from the bluer optical bands (top) to the near-infrared and Wesenheit relations (bottom) highlights the decreasing impact of temperature variations and differential reddening.}\label{fig-pl_feh_type2_jwst}
    \end{figure*}

\begin{figure*}
   \centering
  \includegraphics[trim=0cm 0cm 0.0cm 0cm, clip, width=1.05\linewidth]{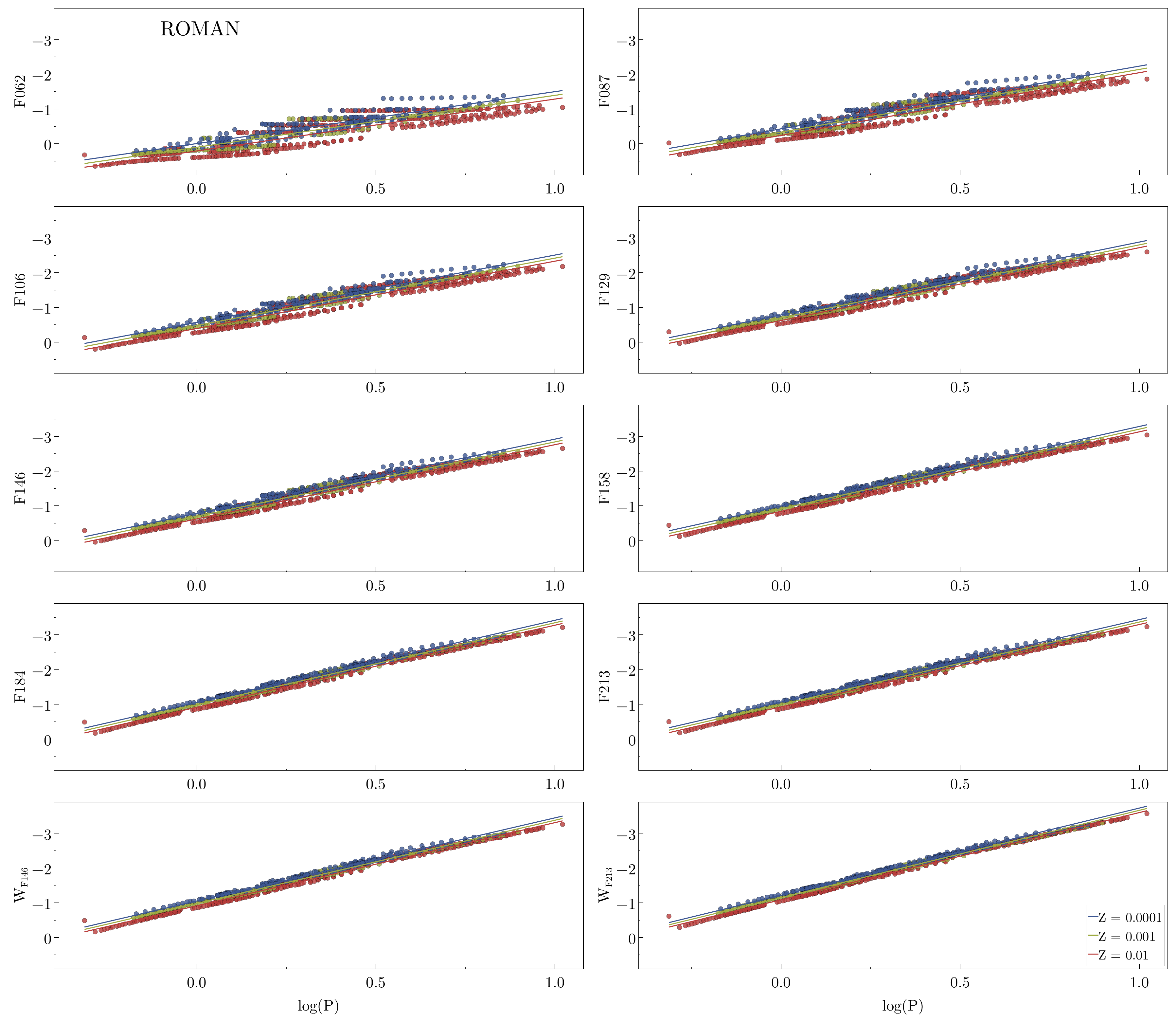}
\caption{Same as Fig.~\ref{fig-pl_feh_type2_jwst} but for the Roman filter system.}\label{fig-pl_feh_type2_roman}
    \end{figure*}

The same relations but for F-mode T2Cs models are shown in Figures \ref{fig-pl_feh_type2_jwst} and \ref{fig-pl_feh_type2_roman}.

In the case of the PW relations, for the JWST photometric system, we defined two reddening-free Wesenheit magnitudes based on the following filter combinations: $\rm W(F150W, F090W-F150W)$ and $\rm W(F277W, F090W-F150W)$. The former adopts the color combination recently employed by \citet{riess_2024ApJ...962L..17R}. The corresponding color term coefficients, derived from the   \citet{cardelli_1989ApJ...345..245C} extinction law, are $\rm R = 0.787$ and $\rm R = 0.291$, respectively, such that:
\begin{equation}
\rm W_{F150W} = F150W - 0.787 \times (F090W - F150W)
\end{equation}
and
\begin{equation}
\rm W_{F277W} = F277W - 0.291 \times (F090W - F150W) .
\end{equation}

Similarly, for the Roman filters, we selected combinations that closely match the JWST passbands in terms of central wavelength: $\rm W(F146, F087-F146)$ and $\rm W(F213, F087-F146)$. The corresponding Wesenheit relations are:
\begin{equation}
\rm W_{F146} = F146 - 0.760 \times (F087 - F146)
\end{equation}
and
\begin{equation}
\rm W_{F213} = F213 - 0.417 \times (F087 - F146) .
\end{equation}
We note that, in particular for RRL models, the plotted PL and PW relations show a non-negligible dependence of the zero point on the adopted metal content. This result was expected on the basis of results already obtained in other photometric systems.

%%%%%%%%%%%%%%%%%%%%%%%%%%%%%%%%%%%%%%%%%%%%%%%
% \begin{figure*}
%   \centering
%   \begin{subfigure}[b]{0.7\linewidth}
%     \centering
%   \includegraphics[trim=0cm 0cm 0.0cm 0cm, clip, width=1.2\linewidth]{pl_feh_type2_jwst.pdf}
%     \caption{xxxx}\label{xxx}
%   \end{subfigure}
  
%   \vspace{0.1cm} % Spaziatura verticale opzionale tra le due righe
  
%   \begin{subfigure}[b]{0.7\linewidth}
%     \centering
%   \includegraphics[trim=0cm 0cm 0.0cm 0cm, clip, width=1.2\linewidth]{pl_feh_type2_jwst.pdf}
%     \caption{JWST photometric system}\label{fig-jwst_lightCurves_example_rrl_F_jwst}
%   \end{subfigure}
% \vspace{-0.1cm}
%   \caption{Theoretical PL and PW relations for Type II Cepheid models in the JWST system. Panels from top to bottom show the relations for the individual filters (F062 through F213) and the Wesenheit magnitudes ($W_{\rm F146}$ and $W_{\rm F213}$, bottom row) as a function of the logarithm of the pulsation period ($\log P$, with $P$ in days). Individual models are color-coded according to their metallicity: $Z = 0.0001$ (blue), $Z = 0.001$ (green), and $Z = 0.01$ (red). Solid lines of corresponding colors represent the best-fit linear relations at each fixed chemical composition. The progressive reduction in dispersion from the bluer optical bands (top) to the near-infrared and Wesenheit relations (bottom) highlights the decreasing impact of temperature variations and differential reddening.}\label{xxxx}
% \end{figure*}
%%%%%%%%%%%%%%%%%%%%%%%%%%%%%%%%%%%%%%%%%%%%%%

In particular, on the basis RRL model predictions, the coefficient of the metallicity term varies from 0.15 to 0.19 mag/dex, whereas for T2Cs the effect is significantly reduced (ranging from 0.06 to 0.1 mag/dex).
These results confirm the possibility, once JWST and Roman photometric data  become available, to use the RRL relations not only to derive the distances of targets with known period and metallicity but also to invert these metal-dependent relations to infer the individual metallicities of RRL with known period and distance \citep[see e.g.][and references therein]{2021MNRAS.508.1064M}.
On the other hand, we confirm that T2C PL and PW relations are almost insensitive to metallicity, even in the JWST and Roman filters, thus making this class of pulsating stars an ideal step in the calibration of a Population II distance scale.
Concerning the slope of the predicted relations, we notice that the values derived for RRL are consistent with the results for T2Cs, again supporting the predicting capabilities of current theoretical scenario in shaping a self consistent calibration of the cosmic distance scale for old stellar populations.

Finally, we note that the slope and the zero point  of the JWST PWZ relations for RRL are consistent within the uncertainties with the preliminary observational results presented in \citet{2026arXiv260221205S}. In the following we will apply the derived theoretical PWZ relations to the RRL sample considered in \citet{2026arXiv260221205S} and compare the inferred distances with the results by these authors and other independent evaluations in the literature.

\subsection{Application to Wolf-Lundmark-Melotte RR Lyrae}
\begin{figure*}
   \centering
  \includegraphics[trim=0cm 22cm 0.0cm 0cm, clip, width=1.0\linewidth]{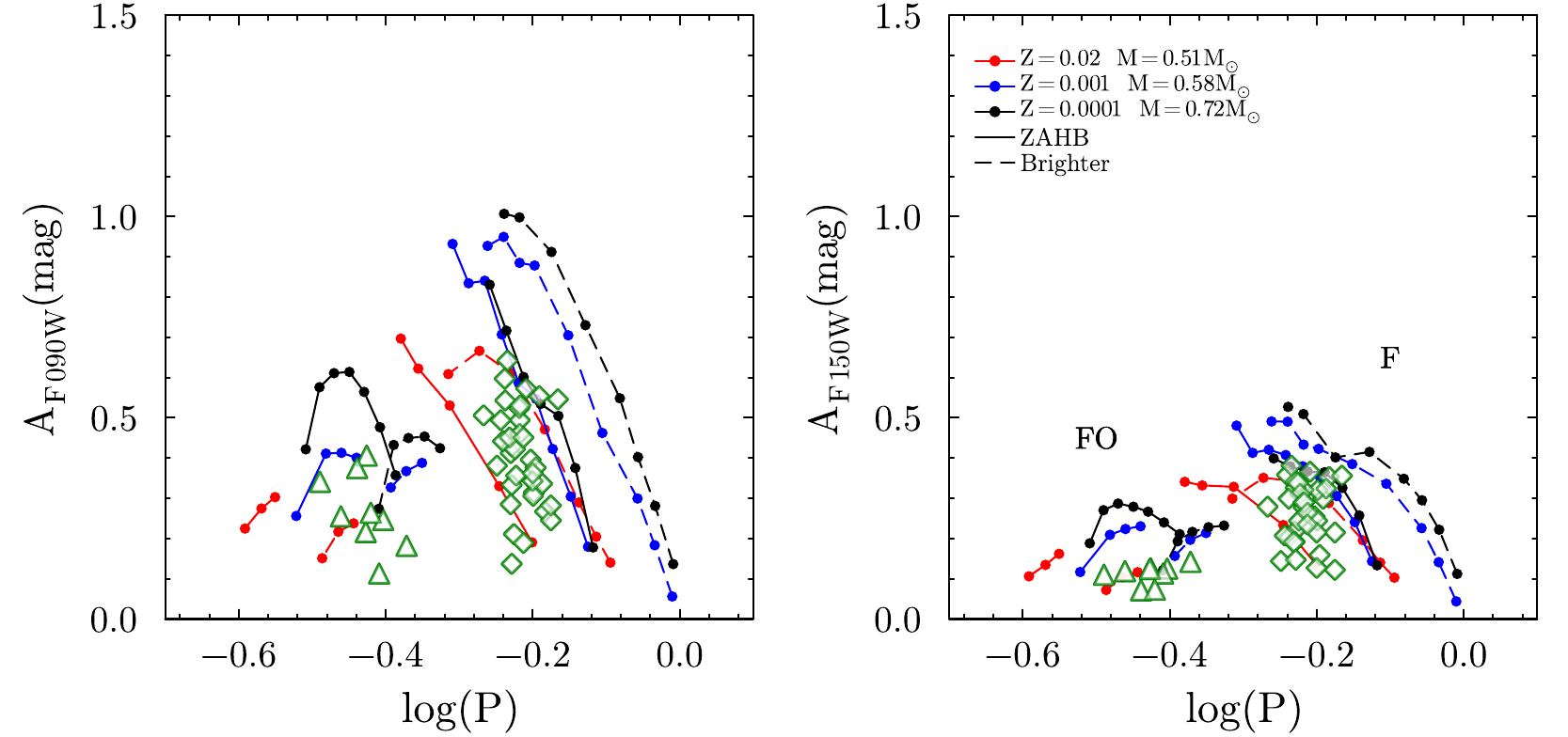}
\caption{Bailey diagram (amplitude versus logarithm of the period) for RR Lyrae variables in the JWST F090W (left panel) and F150W (right panel) filters. The meaning of the curves and symbols for the theoretical models is the same as in Fig. ~\ref{fig-bailey_zy_variable_jwst}. Here, we have added the overlaid green symbols representing the observational data from \citet{2026ApJ..1006..253S}: diamonds indicate fundamental mode pulsators (F), while triangles represent first overtone pulsators (FO).}\label{fig-bailey_zy_variable_rrlyrae_jwst_slaughter2026}
    \end{figure*}

To preliminarily test our theoretical framework, we considered the sample pf RR Lyrae stars observed in the Wolf-Lundmark-Melotte (WLM) dwarf galaxy. We adopted the JWST NIRCam photometric data from \citet{2026ApJ..1006..253S} their Table 9.
Figure \ref{fig-bailey_zy_variable_rrlyrae_jwst_slaughter2026} shows the comparison between the predicted  and the observed (green symbols) Bailey diagram for this RRL sample in the JWST F090W (left panel) and
F150W (right panel) filters. The agreement is quite good even if the adopted models seem to overestimate the pulsation amplitudes and slightly higher mixing length parameter would improve the match of the observed amplitudes.

We then applied the newly derived JWST PWZ relations to a sample of RR Lyrae stars observed in the Wolf-Lundmark-Melotte (WLM) dwarf galaxy. We adopted the JWST NIRCam photometric data from \citet{2026ApJ..1006..253S} their Table 9. For the metallicity, we assumed a mean value of $\mathrm{[Fe/H]} = -1.27$ dex \citep{2009ApJ...699....1L}. Since the available observations were acquired in the F090W and F150W bands, we utilized our derived Wesenheit formulation, $W(F150W, F090W-F150W)$. 
To accurately estimate the intrinsic distance modulus of WLM while accounting for observational uncertainties, we adopted a formal analytical error propagation. For each individual RRL star, the uncertainty on the distance modulus was computed by propagating the measurement errors on the pulsation period and mean magnitudes through the linear Period-Wesenheit-Metallicity (PWZ) relations. Additionally, the intrinsic scatter of the theoretical relations was added in quadrature to the observational error budget of each star. The final distance to the WLM galaxy was then calculated as the mean of the individual distance moduli, with the overall uncertainty derived from the combined variance of the individual measurements divided by the sample size (i.e., the standard error of the mean).
% To accurately estimate the intrinsic distance modulus of WLM while accounting for observational uncertainties, we used the Monte Carlo error propagation method. For each of the 10 RRL stars, we generated 10,000 random realizations of the pulsation period and mean magnitudes, drawing from normal distributions based on their respective measurement errors. Additionally, the intrinsic scatter of the adopted PWZ relations was included as a random perturbation in the calculations. For each individual star, the distance modulus and its uncertainty were calculated as the mean and standard deviation of the simulated distribution. The final distance to the WLM galaxy was then computed as the mean of the individual distance moduli, with the final uncertainty represented by the standard error of the mean.

By applying the relations for F and FO pulsators separately - anchored to our standard solar metallicity of $Z_{\odot} = 0.02$ - we obtained a distance modulus of %$\mu_0 = 24.849 \pm 0.031$ mag
$\mu_0 = 24.855 \pm 0.018$ mag. Alternatively, by fundamentalizing the period of the FO pulsators (as detailed above) and applying our global F+FO relation, we derived a value of %$\mu_0 = 24.845 \pm 0.031$ mag.
$ \mu_0 = 24.851 \pm 0.018$ mag. 

To further investigate the systematic uncertainties of our theoretical framework, we evaluated the impact of the assumed solar metallicity on the calibration. We recalculated the theoretical relations adopting the more recent value of $Z_{\odot} = 0.0139$ (Asplund, Amarsi \& Grevesse 2021). As expected, this variation exclusively affects the zero-point of the theoretical relations, resulting in a median zero-point offset of $-0.027$ mag. Applying this updated calibration to our WLM dataset yields revised distance moduli of $\mu_0 = 24.869 \pm 0.018$ mag (mode-specific) and $\mu_0 = 24.879 \pm 0.018$ mag (global F+FO).

\begin{figure}
   \centering
  \includegraphics[trim=0cm 0cm 0.0cm 0cm, clip, width=1.0\linewidth]{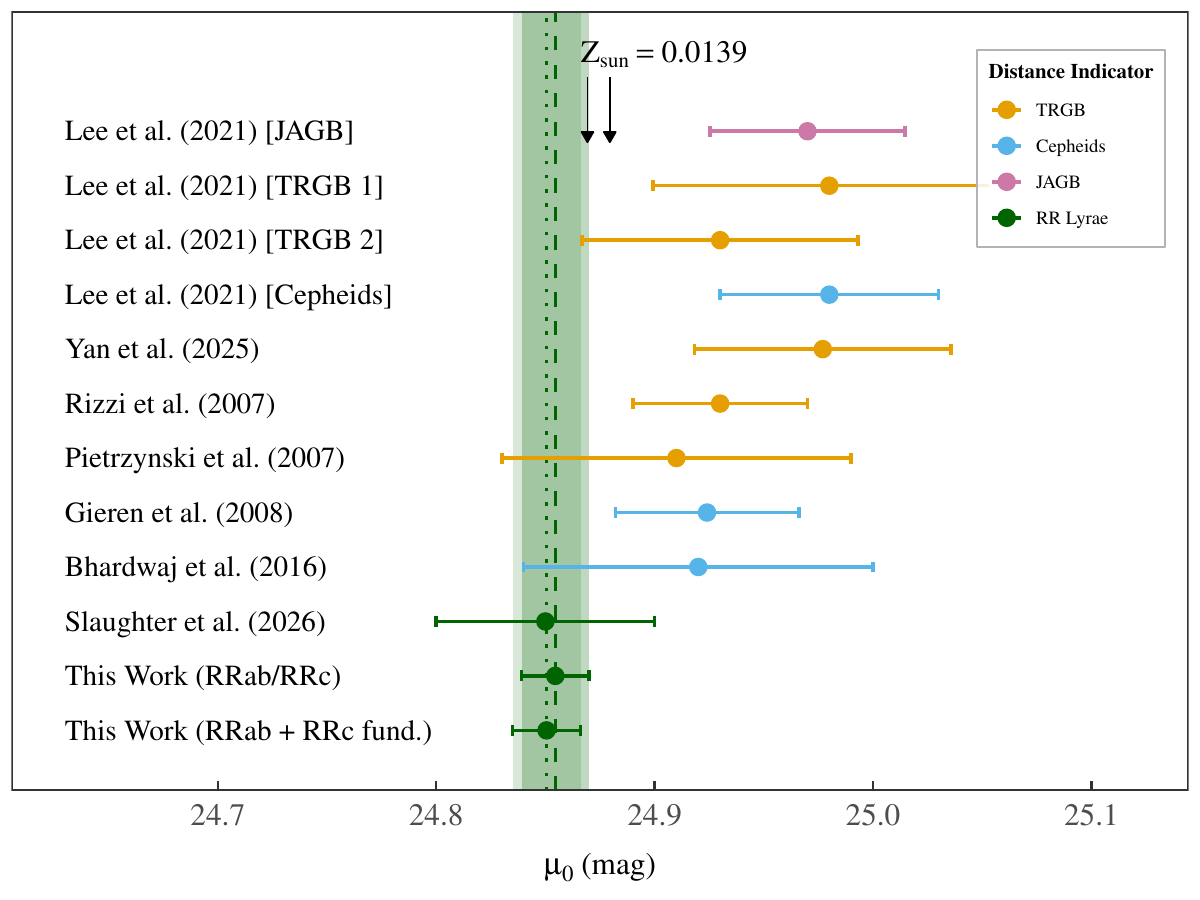}
\caption{Comparison of the intrinsic distance modulus to the WLM galaxy. The plot illustrates the distance moduli obtained in this work using the new theoretical JWST Period-Wesenheit-Metallicity (PWZ) relations in the $\rm W(F150W, F090W-F150W)$ band. The main results, anchored to a standard solar metallicity of $\rm Z_\odot = 0.02$, are presented for two different approaches: adopting mode-specific relations for fundamental (F) and first-overtone (FO) pulsators, and using a global relation with fundamentalized FO variables. The dotted green line indicates the distance obtained with the global relation, while the dashed green line refers to the distance obtained with the mode-specific relations. The shaded green bands represent the uncertainties associated with our new RR Lyrae calibrations. The black arrows at the top of the panel indicate the systematic shift in our derived distances when adopting the more recent solar metallicity value of $\rm Z_\odot = 0.0139$ \citep{2021A&A...653A.141A}.  These new distance estimates are compared against independent indicators from the literature, specifically the Tip of the Red Giant Branch (TRGB), Classical Cepheids, and the J-region Asymptotic Giant Branch (JAGB). Literature measurements include studies by \citet{1993npsp.conf...92L, 2007ApJ...661..815R, 2007AJ....134..594P, 2008ApJ...683..611G, 2016AJ....151...88B, lee_2021ApJ...907..112L, yan_2025ApJ...986...54Y}, alongside with the recent result by \citet{2026arXiv260221205S}}\label{fig-wlm_distance}
    \end{figure}

These results are compared with a selection of independent literature estimates in Figure~\ref{fig-wlm_distance}. As illustrated, our default JWST RR Lyrae-based distances perfectly agree with the determination by \citet{2026arXiv260221205S}. The systematic shift introduced by the $Z_{\odot} = 0.0139$ calibration is indicated by the black arrows at the top of the panel, demonstrating how these revised estimates bring our distances into tighter agreement with the bulk of previous determinations based on Classical Cepheids, the Tip of the Red Giant Branch (TRGB), and the  Asymptotic Giant Branch (JAGB), which generally cluster around $\mu_0 \sim 24.92$ mag. 

Although our estimates remain slightly shorter than the literature average, they are consistent with the measurement by \citet{lee_2021ApJ...907..112L}, \citet{2007AJ....134..594P}, \citet{2016AJ....151...88B} and agree within more than 1$\sigma$ with the other  plotted estimates. It is worth emphasizing that our analysis relies on a specific subsample of observed RR Lyrae stars. Taking into account our predicted metallicity term coefficient, the agreement would significantly improve if the adopted metallicity for this particular subsample were $\sim 0.3$ dex more metal-poor. Such a deviation is perfectly consistent with the intrinsic metallicity spread of the galaxy; indeed, \citet{2009ApJ...699....1L} reported a mean metallicity of $\mathrm{[Fe/H]} = -1.27 \pm 0.04$ dex with a substantial standard deviation of $\sigma = 0.37$ dex. Therefore, if the specific stars in our subsample turned out to be slightly more metal-poor than the galaxy's average, this would fully account for the offset in our distance estimate. This demonstrates that precise distance measurements depend heavily on knowing the true metallicity of the specific stars being observed.

\section{Conclusions}

The nonlinear convective pulsation models presented in this study establish a comprehensive theoretical framework for RRL and T2Cs within the specific filter systems of the JWST and Roman.  
A detailed analysis of the modeled light curves and period-amplitude diagrams shows that pulsation amplitudes decrease progressively toward longer wavelengths, yielding, as expected,  smoother variations in the infrared filters compared to traditional optical bands. Furthermore, we confirm  that the period-amplitude diagrams for both RRL and T2C models reveal a systematic decrease in both pulsation periods and amplitudes as the overall metallicity increases.  Model predictions were also investigated in various color-color planes. In the RRL case, specific band combinations—most notably the $(F087-F184)$ versus $(F184-F213)$ plane in the Roman filter system—demonstrate a striking separation between sequences of different metallicity levels, even though the total spread along the X-axis covers only a few hundredths of a magnitude. Concurrently, across almost all JWST and Roman color-color plots, the slope of the interstellar reddening vector differs from the slope of the calculated model sequences. This geometric decoupling offers a powerful diagnostic space to constrain and infer individual reddening values for observed pulsators by directly comparing empirical data with theoretical predictions.  
The newly derived PL and PW relations demonstrate that while RRL pulsators possess a significant metallicity dependence ranging from 0.15 to $0.19\text{ mag/dex}$ in their zero points, T2C relations remain almost insensitive to metal abundance, with the effect reduced to 0.06 to $0.1\text{ mag/dex}$. This stability positions T2Cs as ideal, only slightly metallicity-dependent standard candles for calibrating the Population II cosmic distance scale. Conversely, the stronger dependence found in RRLs allows for an "inverse" application: when distances and periods are known, these theoretical PLZ and PWZ relations can be inverted to infer individual stellar metallicities. 
 Good agreement is found with the coefficients (which remain somewhat uncertain) of the empirical PWZ relations for RRL in the JWST bands given by \citet{2026arXiv260221205S}. Moreover, the consistency found between the predicted slopes of the RRL and T2C relations reinforces the reliability of the current theoretical framework, providing an excellent benchmark to validate and refine the cosmic distance scale. It is also worth noticing that NIR and MIR PL and PLZ relations can be reliably used for RR Lyrae stars exhibiting the Blazhko effect \citep[see e.g.][and references therein]{jurcsik18}, thus expanding the observational samples that can be adopted to derive the distances of host systems.
As a preliminary application of the developed theoretical scenario, we considered the sample of 44 RRL with JWST observations in WLM recently published by \citet{2026ApJ..1006..253S} and derived independent individual and mean distance values by applying the theoretical PWZ relation in the F090W, F150W filters, predicted in this paper. The obtained values agree with the result by \citet{2026arXiv260221205S} and are consistent within the uncertainties with most of the literature values, still yielding  a slightly shorter distance modulus compared to the bulk of previous determinations based on Classical Cepheids and the TRGB method.
However,  taking into account our predicted metallicity term coefficient, the agreement would significantly improve if the adopted RRL subsample were $\sim 0.3$ dex more metal-poor than the adopted average values, a difference perfectly consistent with the intrinsic metallicity spread of the galaxy \citep{2009ApJ...699....1L}. 
This preliminary test confirms the predictive capability of the presented theoretical scenario.
Looking forward, these theoretical models, relations, and diagnostic patterns are primed for immediate application against incoming high-cadence, deep-sky time-series databases from JWST and Roman. This synergy will drastically extend our reach to variable stars in distant, heavily obscured environments across the Local Universe, allowing for precise distance determinations and the derivation of individual stellar parameters.

\begin{acknowledgements}
We are grateful to our anonymous referee for their useful comments that helped improving the manuscript.
 We acknowledge funding from: INAF GO-GTO grant 2023 “C-MetaLL - Cepheid metallicity in the Leavitt law” (P.I. V. Ripepi); PRIN MUR 2022 project (code 2022ARWP9C) 'Early Formation and Evolution of Bulge and Halo (EFEBHO),' PI: Marconi, M., funded by the European Union – Next Generation EU; Large Grant INAF 2023 MOVIE (P.I. M. Marconi).
T.S. thanks the Istituto Nazionale di Fisica Nucleare (INFN), Naples section, for support through specific initiative QGSKY. Finally, this paper is based on work supported by COST Action CA21136, "Addressing Observational Tensions in Cosmology with Systematics and Fundamental Physics (CosmoVerse)," funded by COST (European Cooperation in Science and Technology). 
AB thanks the funding from the Anusandhan National Research Foundation (ANRF) under the Prime Minister Early Career Research Grant scheme (ANRF/ECRG/2024/000675/PMS). This research was supported by the International Space Science Institute (ISSI) in Bern/Beijing through ISSI/ISSI-BJ International Team project ID $\#$24-603 – ``EXPANDING Universe'' (EXploiting Precision AstroNomical Distance INdicators in the Gaia Universe). 
GDS acknowledges funding from Gaia DPAC through INAF and ASI (2025-10-HH.0; PI: M. G. Lattanzi) and from INFN, Naples Section, through the QGSKY and Moonlight2 initiatives.

\end{acknowledgements}

\bibliographystyle{aa}
\bibliography{paper.bib}

\begin{appendix} 
\section{FO-mode model light curves in the Roman and JWST bands}
\begin{figure*}[h]
  \centering
  \begin{subfigure}{0.77\linewidth}
    \centering
  \includegraphics[trim=0cm 3.0cm 0.0cm 0cm, clip, width=1.0\linewidth]{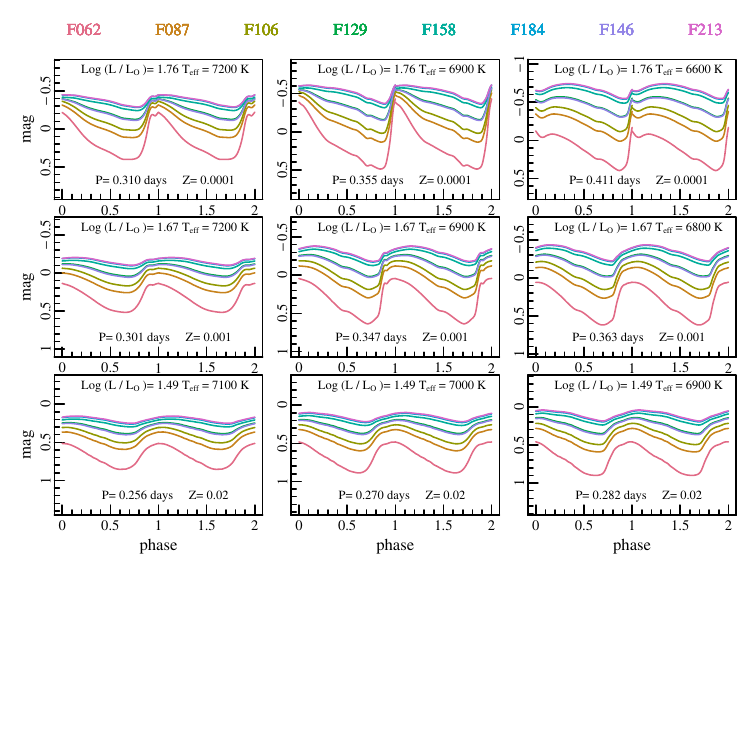}
\caption{Roman photometric system}\label{fig-roman_lightCurves_example_rrl_FO_roman}%
  \end{subfigure}
   
  \begin{subfigure}{0.77\linewidth}
    \centering
  \includegraphics[trim=0cm 3.0cm 0.0cm 0cm, clip, width=1.0\linewidth]{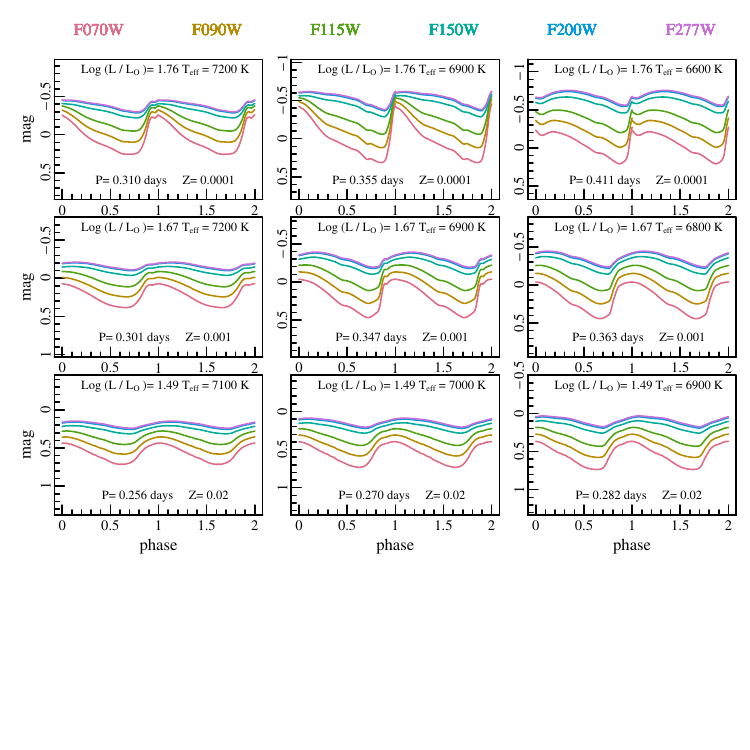}
\caption{JWST photometric system}\label{fig-jwst_lightCurves_example_rrl_FO_jwst}%
  \end{subfigure}
\vspace{-0.1cm}
  \caption{Same as Fig.~\ref{fig-lightCurves_combined}, but for FO-mode pulsators.}\label{figFO-lightCurves_combined}
\end{figure*}

In Figure \ref{figFO-lightCurves_combined} we show an example of predicted light curves for FO-mode RRL models in the Roman (top) and JWST (bottom) bands. The same trends noted for F-mode RRL models are found, apart from the well known and expected more regular/sinusoidal shape.

\section{Predicted period-amplitude diagrams in the JWST bands}

\begin{figure*}[h]
   \centering
  \includegraphics[trim=0cm 0cm 0.0cm 0cm, clip, width=0.7\linewidth]{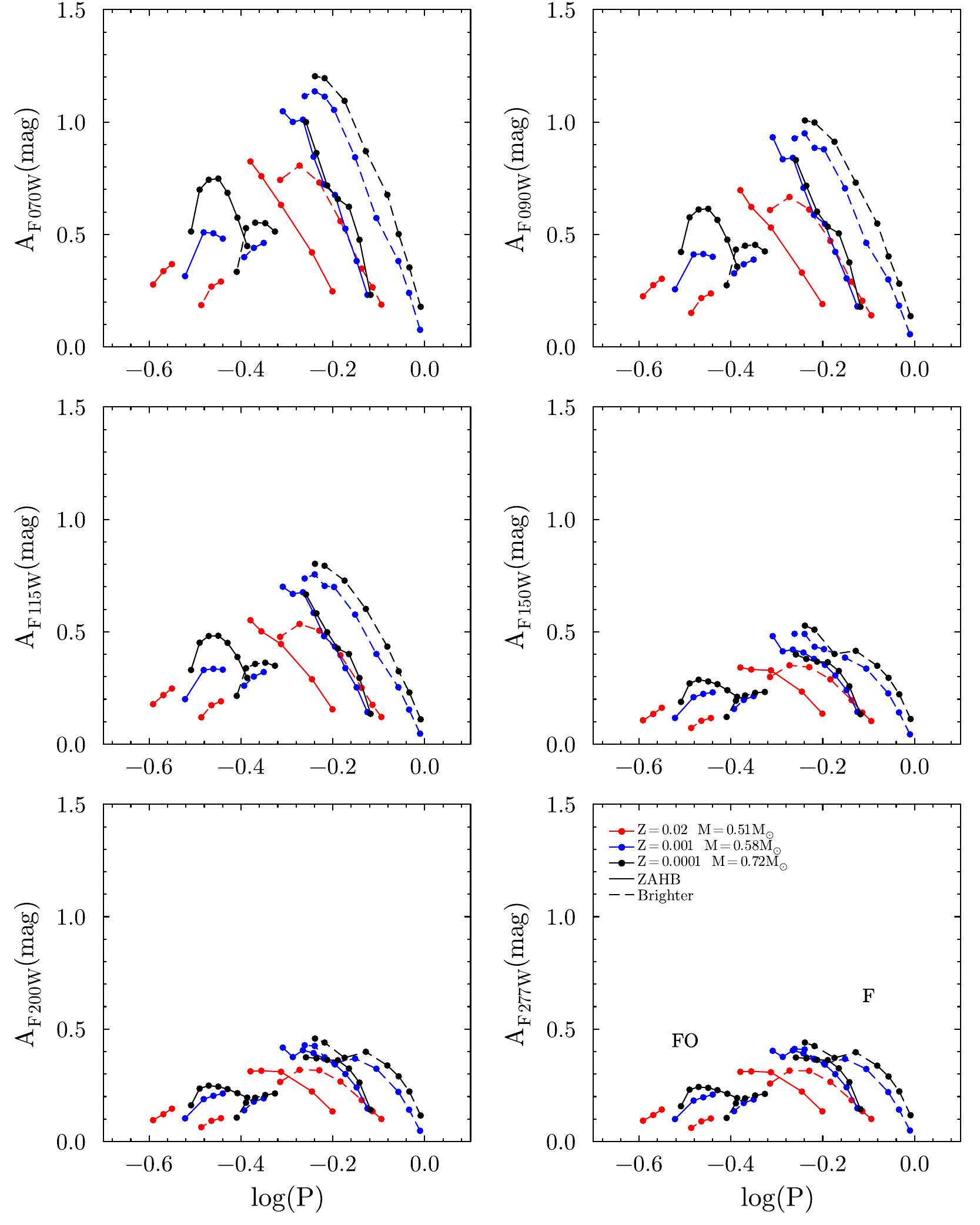}
\caption{The same as Figure \ref{fig-bailey_zy_variable_rrlyrae_roman} but in the JWST filters.}\label{fig-bailey_zy_variable_jwst}%
    \end{figure*}

Figure \ref{fig-bailey_zy_variable_jwst} shows the predicted period-amplitude diagram for RRL in the JWST bands with the same symbols as in Figure \ref{fig-bailey_zy_variable_rrlyrae_roman}.

\begin{figure*}[h]
   \centering
  \includegraphics[trim=0cm 0cm 0.0cm 0cm, clip, width=0.7\linewidth]{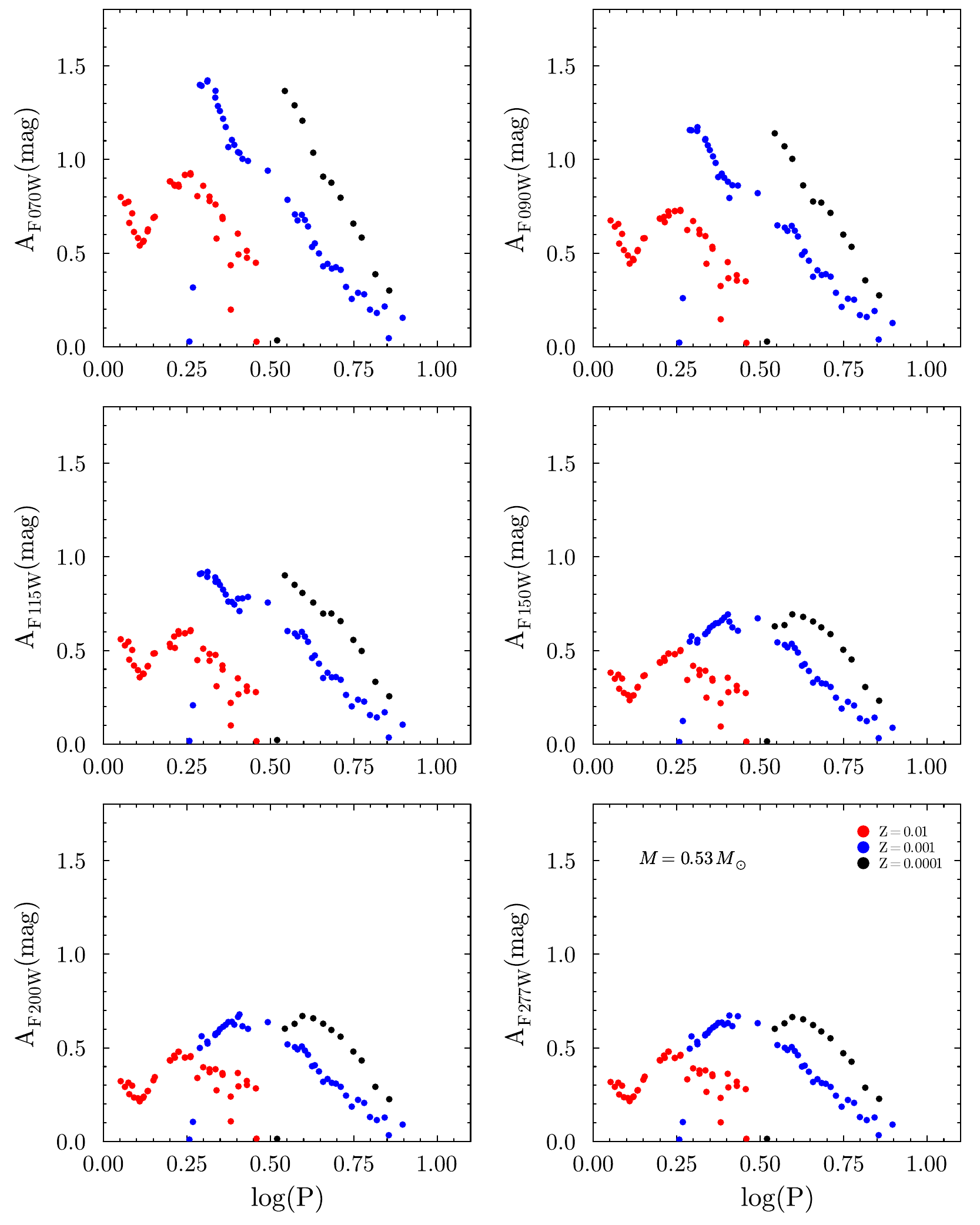}
\caption{The same as in the previous figure but for T2C models}\label{fig-bailey_zy_variable_typeII_jwst_M0.53}%
    \end{figure*}

A similar plot for T2C models is shown in Figure \ref{fig-bailey_zy_variable_typeII_jwst_M0.53}.

\section{Coefficients of the predicted PLZ and PWZ relations based on RRL and T2C models}

In Tables \ref{tab:rrlyrae_relations}-\ref{tab:roman_type2_relations} we report the coefficients of the predicted metal dependent PL and PW relations for RRL and T2C pulsators both in the JWST and Roman photometric bands.

\begin{table*}[htbp]
\centering
\caption{Theoretical Period-Luminosity-Metallicity ($PLZ$) and Period-Wesenheit-Metallicity ($PWZ$) relations for RR Lyrae stars in the JWST filter system: $M = \text{zp} + a \log P + b \text{[Fe/H]}$.}
\label{tab:rrlyrae_relations}
\small
\begin{tabular}{l l c r@{\,$\pm$\,}l r@{\,$\pm$\,}l r@{\,$\pm$\,}l c c c}
\hline\hline
Type & Band / Wesenheit & Mode & \multicolumn{2}{c}{zp} & \multicolumn{2}{c}{$a$ ($\log P$)} & \multicolumn{2}{c}{$b$ ([Fe/H])} & $N$ & rms & $R^2$ \\
\hline
\noalign{\smallskip}
%\multicolumn{12}{c}{\textit{Fundamental Mode (F)}} \\
\noalign{\smallskip}
RRLyrae & F070W & F & -0.076 & 0.020 & -1.523 & 0.072 & 0.162 & 0.012 & 155 & 0.107 & 0.832 \\
RRLyrae & F090W & F & -0.294 & 0.016 & -1.731 & 0.058 & 0.167 & 0.010 & 155 & 0.087 & 0.902 \\
RRLyrae & F115W & F & -0.487 & 0.013 & -1.904 & 0.045 & 0.171 & 0.008 & 155 & 0.068 & 0.946 \\
RRLyrae & F150W & F & -0.739 & 0.008 & -2.167 & 0.027 & 0.181 & 0.005 & 155 & 0.041 & 0.984 \\
RRLyrae & F200W & F & -0.834 & 0.007 & -2.247 & 0.023 & 0.183 & 0.004 & 155 & 0.035 & 0.989 \\
RRLyrae & F277W & F & -0.858 & 0.007 & -2.265 & 0.023 & 0.181 & 0.004 & 155 & 0.035 & 0.989 \\
RRLyrae & $W(F150W, F090W-F150W)$ & F & -1.090 & 0.005 & -2.510 & 0.019 & 0.192 & 0.003 & 155 & 0.028 & 0.994 \\
RRLyrae & $W(F277W, F090W-F150W)$ & F & -0.988 & 0.005 & -2.392 & 0.018 & 0.186 & 0.003 & 155 & 0.028 & 0.994 \\
\hline
\noalign{\smallskip}
%\multicolumn{12}{c}{\textit{First Overtone Mode (FO)}} \\
\noalign{\smallskip}
RRLyrae & F070W & FO & -0.540 & 0.058 & -1.875 & 0.122 & 0.147 & 0.011 & 63 & 0.063 & 0.900 \\
RRLyrae & F090W & FO & -0.775 & 0.048 & -2.097 & 0.102 & 0.150 & 0.009 & 63 & 0.052 & 0.938 \\
RRLyrae & F115W & FO & -0.976 & 0.040 & -2.267 & 0.083 & 0.152 & 0.007 & 63 & 0.042 & 0.962 \\
RRLyrae & F150W & FO & -1.243 & 0.025 & -2.550 & 0.052 & 0.156 & 0.005 & 63 & 0.027 & 0.987 \\
RRLyrae & F200W & FO & -1.342 & 0.022 & -2.631 & 0.046 & 0.157 & 0.004 & 63 & 0.023 & 0.990 \\
RRLyrae & F277W & FO & -1.370 & 0.022 & -2.658 & 0.045 & 0.155 & 0.004 & 63 & 0.023 & 0.991 \\
RRLyrae & $W(F150W, F090W-F150W)$ & FO & -1.612 & 0.014 & -2.907 & 0.030 & 0.161 & 0.003 & 63 & 0.015 & 0.996 \\
RRLyrae & $W(F277W, F090W-F150W)$ & FO & -1.507 & 0.017 & -2.790 & 0.036 & 0.157 & 0.003 & 63 & 0.018 & 0.995 \\
\hline
\noalign{\smallskip}
%\multicolumn{12}{c}{\textit{Fundamental + Fundamentalized First Overtone (F+FO)}} \\
\noalign{\smallskip}
RRLyrae & F070W & F+FO & -0.067 & 0.017 & -1.348 & 0.053 & 0.168 & 0.010 & 218 & 0.105 & 0.830 \\
RRLyrae & F090W & F+FO & -0.290 & 0.014 & -1.594 & 0.043 & 0.170 & 0.008 & 218 & 0.085 & 0.903 \\
RRLyrae & F115W & F+FO & -0.485 & 0.011 & -1.798 & 0.034 & 0.173 & 0.006 & 218 & 0.068 & 0.946 \\
RRLyrae & F150W & F+FO & -0.746 & 0.007 & -2.130 & 0.021 & 0.178 & 0.004 & 218 & 0.042 & 0.984 \\
RRLyrae & F200W & F+FO & -0.843 & 0.006 & -2.228 & 0.019 & 0.179 & 0.003 & 218 & 0.037 & 0.989 \\
RRLyrae & F277W & F+FO & -0.867 & 0.006 & -2.251 & 0.018 & 0.177 & 0.003 & 218 & 0.036 & 0.989 \\
RRLyrae & $W(F150W, F090W-F150W)$ & F+FO & -1.105 & 0.005 & -2.553 & 0.015 & 0.184 & 0.003 & 218 & 0.029 & 0.994 \\
RRLyrae & $W(F277W, F090W-F150W)$ & F+FO & -1.000 & 0.005 & -2.407 & 0.015 & 0.179 & 0.003 & 218 & 0.030 & 0.993 \\
\hline\hline
\end{tabular}
\end{table*}

\begin{table*}
\caption{Best-fit coefficients for the Period-Luminosity-Metallicity and Period-Wesenheit-Metallicity relations of RR Lyrae stars in the Nancy Grace Roman Space Telescope filter system.}
\label{tab:roman_rrlyrae_relations_all}
\centering
\begin{tabular}{ll c r@{ $\pm$ }l r@{ $\pm$ }l r@{ $\pm$ }l c c c}
\hline\hline
\noalign{\smallskip}
Type & Band & Mode & \multicolumn{2}{c}{$\alpha$ (zp)} & \multicolumn{2}{c}{$\beta$ ($\log P$)} & \multicolumn{2}{c}{$\gamma$ ([Fe/H])} & $N_{\rm stars}$ & rms & $R^2$ \\
\noalign{\smallskip}
\hline
\noalign{\smallskip}
RRLyrae & F062 & F & $ 0.123$ & 0.024 & $-1.349$ & 0.085 & 0.167 & 0.014 & 155 & 0.127 & 0.752 \\
RRLyrae & F087 & F & $-0.273$ & 0.017 & $-1.712$ & 0.059 & 0.167 & 0.010 & 155 & 0.088 & 0.898 \\
RRLyrae & F106 & F & $-0.403$ & 0.015 & $-1.825$ & 0.051 & 0.169 & 0.009 & 155 & 0.077 & 0.927 \\
RRLyrae & F129 & F & $-0.596$ & 0.011 & $-2.008$ & 0.038 & 0.174 & 0.006 & 155 & 0.056 & 0.965 \\
RRLyrae & F146 & F & $-0.588$ & 0.011 & $-2.021$ & 0.037 & 0.176 & 0.006 & 155 & 0.056 & 0.966 \\
RRLyrae & F158 & F & $-0.771$ & 0.007 & $-2.198$ & 0.025 & 0.183 & 0.004 & 155 & 0.038 & 0.986 \\
RRLyrae & F184 & F & $-0.828$ & 0.007 & $-2.243$ & 0.024 & 0.184 & 0.004 & 155 & 0.035 & 0.988 \\
RRLyrae & F213 & F & $-0.840$ & 0.007 & $-2.250$ & 0.023 & 0.182 & 0.004 & 155 & 0.035 & 0.989 \\
RRLyrae & $W(F146, F087-F146)$ & F & $-0.828$ & 0.007 & $-2.255$ & 0.023 & 0.182 & 0.004 & 155 & 0.035 & 0.989 \\
RRLyrae & $W(F213, F087-F146)$ & F & $-0.972$ & 0.005 & $-2.378$ & 0.019 & 0.186 & 0.003 & 155 & 0.028 & 0.993 \\
\noalign{\smallskip}
\hline
\noalign{\smallskip}
RRLyrae & F062 & FO & $-0.328$ & 0.068 & $-1.694$ & 0.143 & 0.153 & 0.013 &  63 & 0.073 & 0.861 \\
RRLyrae & F087 & FO & $-0.753$ & 0.049 & $-2.078$ & 0.103 & 0.150 & 0.009 &  63 & 0.053 & 0.935 \\
RRLyrae & F106 & FO & $-0.886$ & 0.044 & $-2.182$ & 0.093 & 0.151 & 0.008 &  63 & 0.047 & 0.950 \\
RRLyrae & F129 & FO & $-1.092$ & 0.033 & $-2.380$ & 0.070 & 0.153 & 0.006 &  63 & 0.036 & 0.974 \\
RRLyrae & F146 & FO & $-1.081$ & 0.033 & $-2.383$ & 0.070 & 0.153 & 0.006 &  63 & 0.036 & 0.974 \\
RRLyrae & F158 & FO & $-1.276$ & 0.023 & $-2.583$ & 0.049 & 0.157 & 0.004 &  63 & 0.025 & 0.989 \\
RRLyrae & F184 & FO & $-1.335$ & 0.022 & $-2.627$ & 0.046 & 0.157 & 0.004 &  63 & 0.024 & 0.990 \\
RRLyrae & F213 & FO & $-1.348$ & 0.022 & $-2.635$ & 0.046 & 0.156 & 0.004 &  63 & 0.023 & 0.990 \\
RRLyrae & $W(F146, F087-F146)$ & FO & $-1.330$ & 0.023 & $-2.615$ & 0.047 & 0.156 & 0.004 &  63 & 0.024 & 0.990 \\
RRLyrae & $W(F213, F087-F146)$ & FO & $-1.485$ & 0.017 & $-2.762$ & 0.036 & 0.157 & 0.003 &  63 & 0.019 & 0.994 \\
\noalign{\smallskip}
\hline
\noalign{\smallskip}
RRLyrae & F062 & F+FO & $ 0.134$ & 0.020 & $-1.148$ & 0.062 & 0.174 & 0.011 & 218 & 0.123 & 0.748 \\
RRLyrae & F087 & F+FO & $-0.268$ & 0.014 & $-1.572$ & 0.044 & 0.170 & 0.008 & 218 & 0.087 & 0.899 \\
RRLyrae & F106 & F+FO & $-0.399$ & 0.013 & $-1.702$ & 0.039 & 0.172 & 0.007 & 218 & 0.077 & 0.928 \\
RRLyrae & F129 & F+FO & $-0.597$ & 0.010 & $-1.927$ & 0.029 & 0.174 & 0.005 & 218 & 0.057 & 0.965 \\
RRLyrae & F146 & F+FO & $-0.590$ & 0.009 & $-1.943$ & 0.029 & 0.175 & 0.005 & 218 & 0.057 & 0.967 \\
RRLyrae & F158 & F+FO & $-0.778$ & 0.007 & $-2.171$ & 0.020 & 0.179 & 0.004 & 218 & 0.040 & 0.986 \\
RRLyrae & F184 & F+FO & $-0.836$ & 0.006 & $-2.224$ & 0.019 & 0.180 & 0.003 & 218 & 0.037 & 0.988 \\
RRLyrae & F213 & F+FO & $-0.849$ & 0.006 & $-2.232$ & 0.019 & 0.178 & 0.003 & 218 & 0.037 & 0.989 \\
RRLyrae & $W(F146, F087-F146)$ & F+FO & $-0.835$ & 0.006 & $-2.225$ & 0.019 & 0.179 & 0.003 & 218 & 0.037 & 0.988 \\
RRLyrae & $W(F213, F087-F146)$ & F+FO & $-0.983$ & 0.005 & $-2.387$ & 0.015 & 0.180 & 0.003 & 218 & 0.030 & 0.993 \\
\noalign{\smallskip}
\hline
\end{tabular}
\tablefoot{The relations follow the form: ${\rm Mag} = \alpha + \beta \log P + \gamma {\rm [Fe/H]}$. Pulsation modes are categorized as Fundamental (F), First Overtone (FO), and the global fundamentalized sample (F+FO).}
\end{table*}

\begin{table*}
\caption{Best-fit coefficients for the Period-Luminosity-Metallicity and Period-Wesenheit-Metallicity relations of Type II Cepheids in the JWST filter system.}
\label{tab:jwst_type2_relations}
\centering
\begin{tabular}{l l r@{ $\pm$ }l r@{ $\pm$ }l r@{ $\pm$ }l c c c}
\hline\hline
\noalign{\smallskip}
Type & Band & \multicolumn{2}{c}{$\alpha$ (zp)} & \multicolumn{2}{c}{$\beta$ ($\log P$)} & \multicolumn{2}{c}{$\gamma$ ([Fe/H])} & $N_{\rm stars}$ & rms & $R^2$ \\
\noalign{\smallskip}
\hline
\noalign{\smallskip}
TYPE2 & F070W & $ 0.018$ & 0.013 & $-1.634$ & 0.022 & 0.096 & 0.007 & 734 & 0.173 & 0.880 \\
TYPE2 & F090W & $-0.236$ & 0.011 & $-1.828$ & 0.018 & 0.094 & 0.006 & 734 & 0.142 & 0.932 \\
TYPE2 & F115W & $-0.467$ & 0.008 & $-2.006$ & 0.015 & 0.085 & 0.005 & 734 & 0.113 & 0.963 \\
TYPE2 & F150W & $-0.780$ & 0.006 & $-2.249$ & 0.010 & 0.079 & 0.003 & 734 & 0.075 & 0.987 \\
TYPE2 & F200W & $-0.900$ & 0.004 & $-2.368$ & 0.008 & 0.071 & 0.003 & 734 & 0.061 & 0.992 \\
TYPE2 & F277W & $-0.918$ & 0.005 & $-2.361$ & 0.008 & 0.079 & 0.003 & 734 & 0.061 & 0.992 \\
TYPE2 & $W(F150W, F090W-F150W)$ & $-1.208$ & 0.003 & $-2.581$ & 0.005 & 0.067 & 0.002 & 734 & 0.040 & 0.997 \\
TYPE2 & $W(F277W, F090W-F150W)$ & $-1.076$ & 0.003 & $-2.483$ & 0.006 & 0.074 & 0.002 & 734 & 0.046 & 0.996 \\
\noalign{\smallskip}
\hline
\end{tabular}
\tablefoot{The relations follow the analytical form: ${\rm Mag} = \alpha + \beta \log P + \gamma {\rm [Fe/H]}$.}
\end{table*}

\begin{table*}
\caption{Best-fit coefficients for the Period-Luminosity-Metallicity and Period-Wesenheit-Metallicity relations of Type II Cepheids in the Nancy Grace Roman Space Telescope filter system.}
\label{tab:roman_type2_relations}
\centering
\begin{tabular}{l l r@{ $\pm$ }l r@{ $\pm$ }l r@{ $\pm$ }l c c c}
\hline\hline
\noalign{\smallskip}
Type & Band & \multicolumn{2}{c}{$\alpha$ (zp)} & \multicolumn{2}{c}{$\beta$ ($\log P$)} & \multicolumn{2}{c}{$\gamma$ ([Fe/H])} & $N_{\rm stars}$ & rms & $R^2$ \\
\noalign{\smallskip}
\hline
\noalign{\smallskip}
TYPE2 & F062 & $ 0.242$ & 0.015 & $-1.495$ & 0.026 & 0.106 & 0.008 & 734 & 0.198 & 0.826 \\
TYPE2 & F087 & $-0.211$ & 0.011 & $-1.805$ & 0.019 & 0.094 & 0.006 & 734 & 0.145 & 0.927 \\
TYPE2 & F106 & $-0.366$ & 0.009 & $-1.937$ & 0.016 & 0.088 & 0.005 & 734 & 0.125 & 0.952 \\
TYPE2 & F129 & $-0.598$ & 0.007 & $-2.097$ & 0.013 & 0.081 & 0.004 & 734 & 0.098 & 0.974 \\
TYPE2 & F146 & $-0.598$ & 0.007 & $-2.139$ & 0.012 & 0.080 & 0.004 & 734 & 0.092 & 0.978 \\
TYPE2 & F158 & $-0.820$ & 0.005 & $-2.288$ & 0.009 & 0.077 & 0.003 & 734 & 0.069 & 0.989 \\
TYPE2 & F184 & $-0.894$ & 0.005 & $-2.363$ & 0.008 & 0.071 & 0.003 & 734 & 0.061 & 0.992 \\
TYPE2 & F213 & $-0.906$ & 0.004 & $-2.371$ & 0.008 & 0.071 & 0.003 & 734 & 0.060 & 0.992 \\
TYPE2 & $W(F146, F087-F146)$ & $-0.893$ & 0.004 & $-2.393$ & 0.007 & 0.069 & 0.002 & 734 & 0.056 & 0.993 \\
TYPE2 & $W(F213, F087-F146)$ & $-1.067$ & 0.003 & $-2.510$ & 0.006 & 0.065 & 0.002 & 734 & 0.045 & 0.996 \\
\noalign{\smallskip}
\hline
\end{tabular}
\tablefoot{The relations follow the analytical form: ${\rm Mag} = \alpha + \beta \log P + \gamma {\rm [Fe/H]}$.}
\end{table*}

\section{Predicted global RRL PL and PW relations}
\begin{figure*}
\centering
\includegraphics[trim=0cm 0cm 0.0cm 0cm, clip, width=1.05\linewidth]{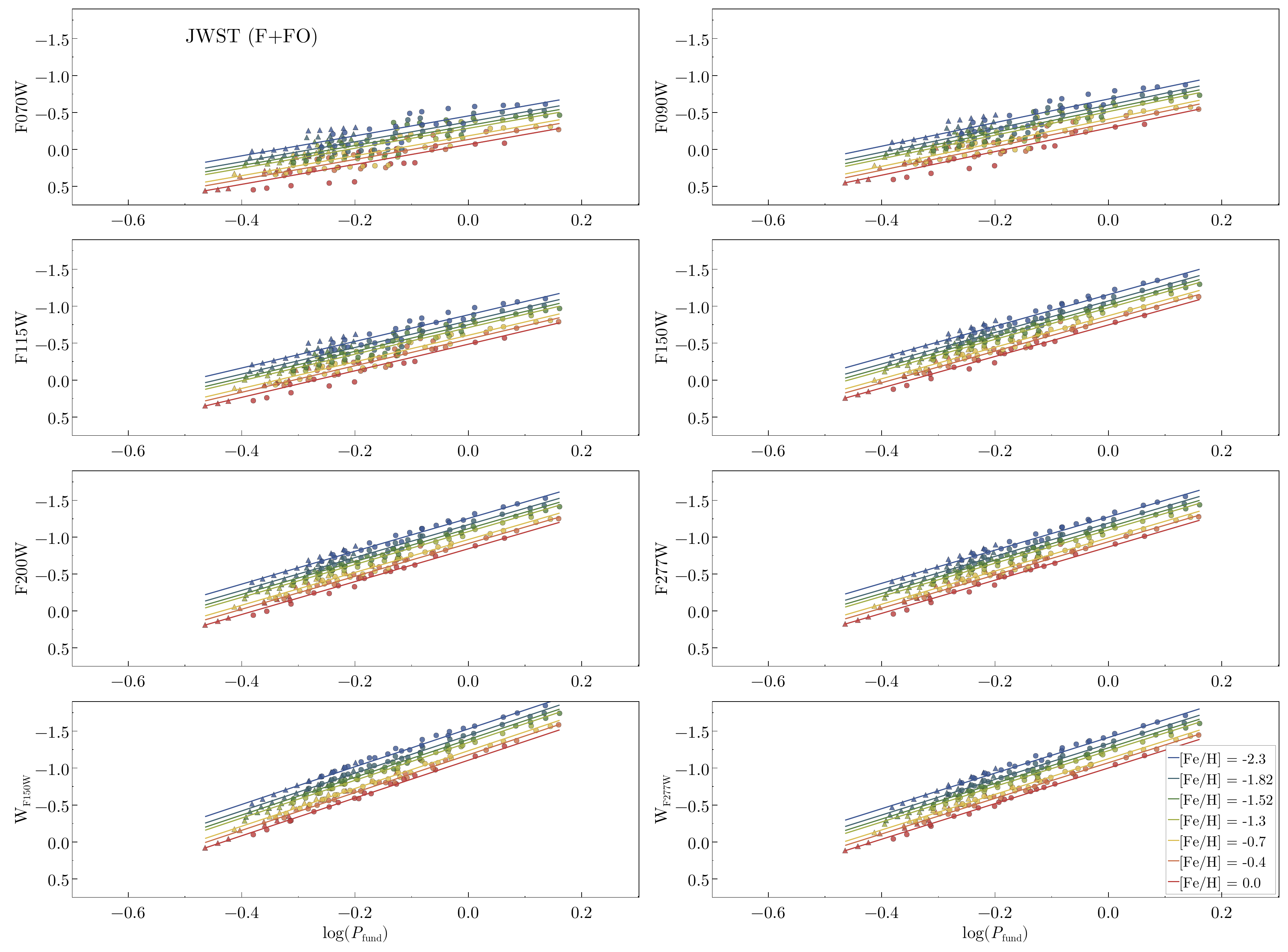}
\caption{Theoretical PLZ and PWZ relations for the combined sample of Fundamental and First Overtone RR Lyrae models in the JWST filter system. The periods of First Overtone pulsators have been fundamentalized using the relation $\rm \log P_{\rm fund} = \log P_{1\rm O} + 0.127$ \citep{marconi2015}. Panels from top to bottom show the relations for the individual passbands and the Wesenheit magnitudes, plotted against the fundamentalized period $\log P_{\rm fund}$. Individual models are color-coded according to their metallicity, ranging from $-2.30$ to $0.00$~dex, as indicated in the legend. Solid lines represent the corresponding best-fit linear relations at each fixed chemical composition.}
\label{fig:pl_feh_fund_rrlyrae_jwst}
\end{figure*}

\begin{figure*}
\centering
\includegraphics[trim=0cm 0cm 0.0cm 0cm, clip, width=1.05\linewidth]{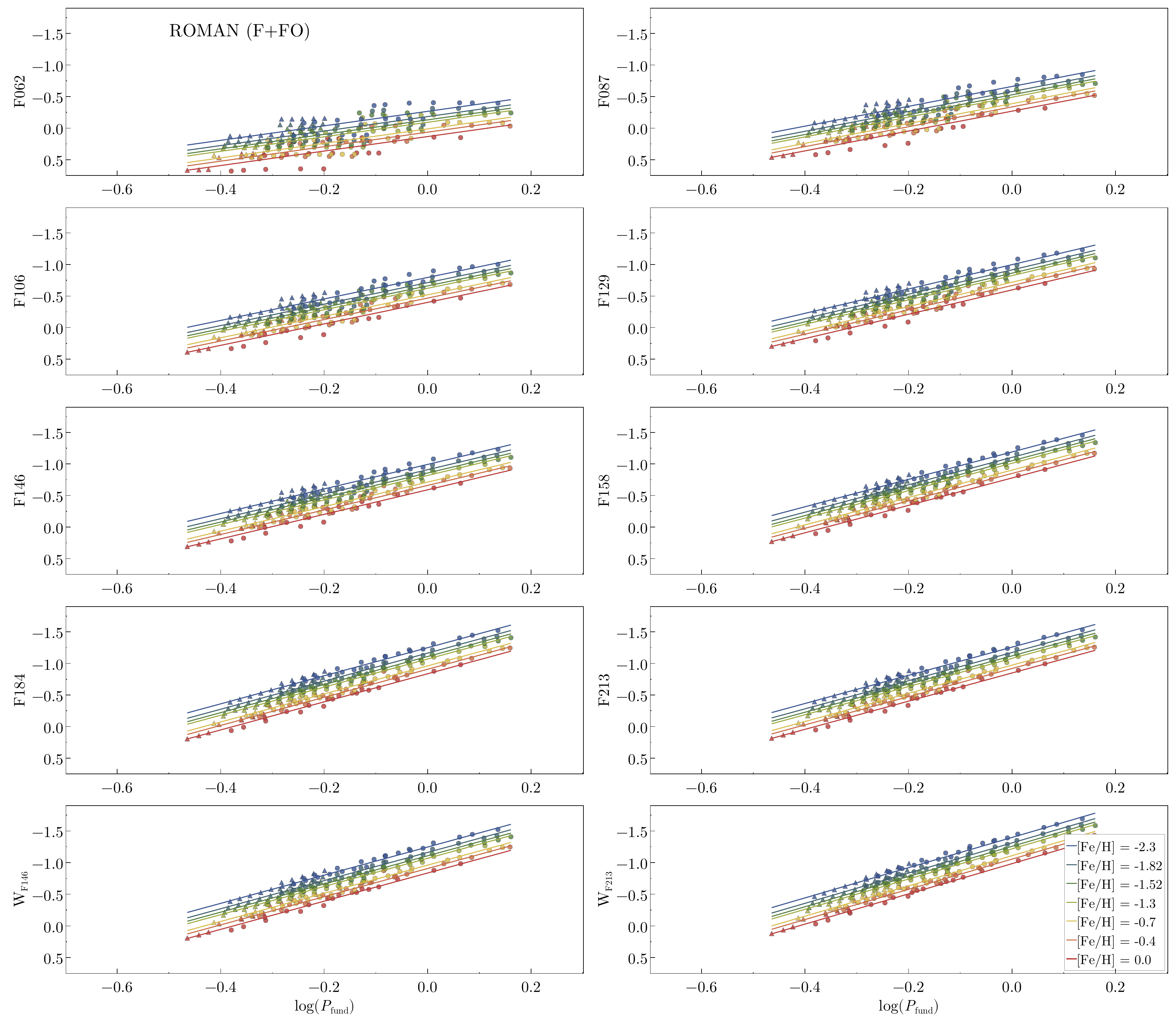}
\caption{Same as Fig.~\ref{fig:pl_feh_fund_rrlyrae_jwst} but for Roman filter system.}
\label{fig:pl_feh_fund_rrlyrae_roman}
\end{figure*}

Figures \ref{fig:pl_feh_fund_rrlyrae_jwst} and \ref{fig:pl_feh_fund_rrlyrae_roman} show the predicted relations obtained by combining F-mode with fundamentalized FO-mode RRL models, in the JWST and Roman bands respectively.

\end{appendix}


\end{document}